\documentclass[format=acmsmall, screen=true, authorversion=true]{acmart}

\usepackage{booktabs} 

\usepackage[ruled]{algorithm2e} 
\usepackage{subfig}
\usepackage{graphics}
\usepackage{amsmath}
\usepackage{wasysym}
\usepackage{amssymb}

\SetAlFnt{\small}
\SetAlCapFnt{\small}
\SetAlCapNameFnt{\small}
\SetAlCapHSkip{0pt}
\IncMargin{-\parindent}

\acmJournal{TIIS}
\acmYear{2020} \acmVolume{10} \acmNumber{2} \acmArticle{3} \acmMonth{6} \acmPrice{15.00}\acmDOI{10.1145/3366501}

\begin{document}

\title[Adaptive Health Coach for Aerobic Exercises]{Designing an AI Health Coach and Studying its Utility in Promoting Regular Aerobic Exercise}
\author{Shiwali Mohan}
\email{shiwali.mohan@parc.com}
\author{Anusha Venkatakrishnan}
\email{anusha.venkatakrishnan@parc.com}
\affiliation{%
  \institution{Palo Alto Research Center}
  \streetaddress{3333 Coyote Hill Road}
  \city{Palo Alto}
  \state{CA}
  \postcode{94306}
  \country{USA}}
\author{Andrea L. Hartzler}
\affiliation{
    \institution{Department of Biomedical Informatics and Medical Education, University of Washington}}

\begin{abstract}
Our research aims to develop interactive, social agents that can coach people to learn new tasks, skills, and habits. In this paper, we focus on coaching sedentary, overweight individuals (i.e., "trainees") to exercise regularly. We employ adaptive goal setting in which the intelligent health coach generates, tracks, and revises personalized exercise goals for a trainee. The goals become incrementally more difficult as the trainee progresses through the training program. Our approach is model-based - the coach maintains a parameterized model of the trainee's aerobic capability that drives its expectation of the trainee's performance. The model is continually revised based on trainee-coach interactions. The coach is embodied in a smartphone application, \textsc{NutriWalking},  which serves as a medium for coach-trainee interaction. We adopt a task-centric evaluation approach for studying the utility of the proposed algorithm in promoting regular aerobic exercise. We show that our approach can adapt the trainee program not only to several trainees with different capabilities, but also to how a trainee's capability improves as they begin to exercise more. Experts rate the goals selected by the coach better than other plausible goals, demonstrating that our approach is consistent with clinical recommendations. Further, in a $6$-week observational study with sedentary participants, we show that the proposed approach helps increase exercise volume performed each week.
\end{abstract}

\maketitle

\section{Introduction}
As artificial intelligence (AI) algorithms and technologies become reliable and robust, they are being employed to solve complex problems in health, education, research, transportation, and other critical contexts. These areas are significantly different from domains such as computer games, abstract problems like the blocks world, curated machine learning and natural language processing data sets that have fueled AI research in past decades. These \emph{real-world} domains require that AI methods be competent at understanding and collaborating with humans - be  \emph{human-aware} \cite{khampapati2018}. Not surprisingly, questions related to relationships between intelligent systems and humans have taken center stage in AIimages research. This paper proposes a \emph{human-aware AI system} that collaborates with humans to support their learning and explores how such systems can be evaluated.

We are interested in designing an interactive agent for one such complex, real-world domain - promoting healthy lifestyles. Unhealthy behaviors are associated with several chronic illnesses such as diabetes and cardiovascular disease. The challenge of developing effective methods for improving health behaviors is becoming critically important as the affected population increases around the world. Medical studies have  explored several strategies - \emph{interventions} - to promote healthier behaviors \cite{kahn2002effectiveness,heath2012evidence}. Examples include disseminating information to change attitudes toward health \cite{huhman2017social}, counseling to learn and maintain healthy habits \cite{lefevre2014behavioral}, and creating supportive social environments \cite{smith2017association}. Most effective of these interventions is individual counseling received in personal meetings \cite{eden2002does,berra2015making} or over telephone \cite{Eakin2007TelephoneReview,goode2012telephone}. Although useful, such counseling is resource-intensive for training counsellors and is difficult to scale to a large population. Consequently, exploring intelligent solutions for providing individual health counselling is a valuable pursuit.

This paper focuses on the design and analysis of a particular type of intelligent solution - an intelligent health coach. The primary role of a coaching agent in a human-agent collaborative setting is to help the human trainee gain knowledge, skills, and tools to perform a new task. The coach may also motivate a trainee to strive for challenging variations of the task and/or provide emotional support in case of continued failures. If intelligent agents are to be successful in coaching a person, they must take into account the person's specific needs, circumstances, and capability in their reasoning and decision making. People vary greatly along these factors and these factors also evolve over time with experience with tasks. It is critical that a coaching agent represent a person's state describing these factors as well as how that state changes over time. The coaching agent must tailor its coaching strategy to each specific person (\textit{personal adaptation}) as well as to how a person evolves while training with a coach (\textit{temporal adaptation}). Training a person for a new task may take a long time ($2-3$ months) and several sessions. A coaching agent, therefore, is required to be a long-living system that maintains an ongoing interaction with its trainee.

The intelligent health coach described here is designed to train overweight, sedentary individuals to develop capability and strength for regular aerobic exercise. Engaging in regular aerobic exercise, such as biking or walking, increases overall energy expenditure above and beyond resting energy expenditure. This helps maintain a healthy weight as well as improve outcomes for weight-related co-morbidities, such as type II Diabetes Mellitus, dyslipidemia \cite{garber2011quantity}, that affect a large number of people around the world. In this context, walking is a preferred form of exercise because it is simple, versatile, requires limited resources, and is easily adaptable to individuals with varying capabilities. However, people need support for selecting specific goals to work on and close monitoring to develop exercising habits. Therefore, a coaching agent for walking has the potential to make this support accessible for a large number of people at low cost.

This paper develops a computational formulation \cite{mohan2017designing} of the goal setting strategy of coaching that is employed in human-human training scenarios \cite{shilts2004goal}. Here, the coach sets relevant and appropriate goals for the trainee. To be effective, a coach must set goals that are \emph{difficult}, yet \emph{attainable}. Goals must induce effort for a trainee to improve performance but should not be too difficult for success. If a trainee walks $15$ minutes every day, a difficult yet attainable goal could be to walk $25$ minutes every day. Additionally, the goals must be specific by providing a clear and narrow target for which the required amount of effort can be estimated. Goals must be proximal and mobilize effort in the near future. Long-term, distal goals make it easy to postpone effort. For example, \emph{Walk for $20$ minutes tomorrow evening} is more effective than \emph{walk more}, which is both general and distal.

The coaching agent is embodied in the \textsc{NutriWalking} smartphone application through which it interacts with a human trainee. During those interactions, the coach assesses the trainee's current state, recommends exercise goals, and evaluates the trainee's performance. To set attainable goals, the coach maintains hypotheses about the trainee's current aerobic capability. It employs a \emph{trainee model} which is continually revised based on how the trainee performs on recommended goals. The coach's recommendations are then heuristically biased by the model's estimation.

This paper takes an important step toward a principled, task-centric efficacy analysis of intelligent behavior-change agents. We deployed this coach to promote aerobic exercise in $21$ adults with co-morbid diabetes and depression. The participants were volunteers from an integrated health system - Kaiser Permanente Washington (formerly Group Health Cooperative). Participants interacted with the coach deployed on their personal smartphones daily for $6$ weeks. Through those interactions, the coach helped them select walking goals that were personalized based on their previous activity levels. The participants monitored and tracked their success or failure at achieving these exercise goals. Daily behavioral data (physical activity self-reports) was collected while trainees engaged in their day-to-day lives rather than in the structured environment of laboratory experiments typical of interactive AI evaluations.

Our computational and empirical findings make the following contributions:

\begin{itemize}
    \item Computational contributions: through domain analysis with an expert, we developed:
    \begin{enumerate}
        \item  a parameterized model for growth in aerobic capability which encodes factors that clinical experts use in their prescription of physical activity.  It affords online revisions by changing parameters as the agent gathers information about the trainee;
        \item a formulation for adaptive goal setting for exercises that can be used by an interactive coaching agent for personal as well as temporal adaptation; and
        \item an implementation of the interactive coach embedded in the \textsc{NutriWalking} smartphone application.
    \end{enumerate}
    \item Empirical contributions: through a task-centric evaluation, we show that:
    \begin{enumerate}
        \item our approach to adaptive goal setting can adapt exercise goals for various types of trainees;
        \item the goals recommended by the coaching agent align with recommendations clinical experts make;
        \item the \textsc{NutriWalking} smartphone application is usable by our target population, who interact with application as expected;
        \item the intelligent health coach sustains adaptive goal setting for $6$ weeks with participants embedded in their daily lives, which makes evaluating long-term behavior change feasible. It promotes aerobic exercise and helps in increasing the volume of aerobic exercise performed each week;
        \item  various design choices to support trainee-coach interactions are usable by human trainees and the information collected using those interactions is useful in personalizing exercises to the trainee.
   \end{enumerate}

\end{itemize}

\section{Related Work}
Use of technology to affect health behavior change \cite{webb2010using} has been gaining popularity as mobile phones and personal computers become more pervasive. A large number of interventions conducted through technological medium are authored by experts and not personalized to a person's individual needs. The role of technology has been limited to delivering the content in a timely and accessible fashion.

Recently researchers have begun to study how adaptive intelligent systems can aid delivery of behavior change interventions. Prior work \citep{schulman2011intelligent} has studied how conversational agents and dialog systems can be used for motivational interviewing (MI) to promote exercise and healthy eating. \citet{schulman2011intelligent} proposed semantics for MI dialog moves and evaluated the resulting conversational interface with $17$ participants in a laboratory experiment. Each participant had $3$ conversations with the agent and was asked to pretend as if a day had passed between conversations. An expert trained in MI counseling rated a subset of conversations by assessing empathy and fidelity to MI. The agent received high ratings suggesting that the proposed semantics were useful for applying MI technique in conversations.

MI has also been explored from the perspective of virtual agent design \cite{lisetti2013can}. This effort focused on designing a human-like virtual agent to help patients with alcohol addiction. The proposed agent is embodied in a virtual persona and the research has explored how the agent can use facial gestures to display affective empathy and can demonstrate verbal reflexive listening by paraphrasing and summarizing the patient's responses. This agent was evaluated online by recruiting $81$ participants and having them randomly interact with the proposed empathic agent, a non-empathic control agent, or a text-based control system. The results showed that the empathic agent was perceived to be more useful and was more enjoyable to interact with (among other metrics) compared with controls.

Others \cite{Tielman2015} have explored using an ontology-based question system for trauma counseling. The focus of their effort was personalizing the content of the questions to each individual user using an ontology of events. In a laboratory study with $24$ participants, they demonstrated that personalized questions resulted in elicitation of more content from participants in comparison to standard questions.

It is worth noting that evaluation in prior work is largely limited to studying various dimensions \emph{interactivity} and \emph{acceptability} of the proposed AI technology. It is largely silent on whether these methods are useful in producing behavior change in human trainees. Evaluating whether an intelligent agent can produce behavior change is challenging and requires long-term deployment, often for several weeks. Prior work in human-computer interaction \cite{konrad2015finding, hollis2015change} demonstrates how long-term deployments in user populations can be used to measure behavior change and evaluate efficacy of interventions. However, the methods studied did not incorporate AI algorithms and did not personalize interventions to each specific individual or adapt to different contexts of human-agent interaction. Recent work \cite{sidner2018creating} has studied the long-term efficacy of AI technology in the form of dialog agents to support isolated older adults. Although the agents described by \cite{sidner2018creating} do not explicitly target health, it is an example of AI agents supporting wellness goals of their human partners.

Recent work \cite{nahum2015building}  proposed an overarching conceptual vision of how technology can be built to support Just In Time Adaptive Interventions (JITAI) for health-related behavior change. Our approach can be considered an exemplar of such a system. The distal outcome of our approach is to transition the trainee from a sedentary lifestyle to the AHA-recommended aerobic exercise volume per week. The proximal outcome is weekly exercise. The decision point is every day and our adaptive algorithms implement several decision rules based on how experts reason about exercise volume prescription.

In order to develop our approach, we propose a trainee model that drives the coach's expectations about the trainee and is useful in picking ideal goals. Previously, the trainee (or learner) models have been studied by the intelligent tutoring systems community \cite{desmarais2012review}. These models assume that what drives a learner's performance is a set of discrete skills that they possess. This discrete representation is not sufficient for representing factors that influence a trainee's performance on exercises such as walking. These models usually represent beliefs about the learner's cognitive skills, such as addition or multiplication. This is not sufficient for coaching exercises. Even if a trainee knows how to walk, they can have substantially different performance on walking for $15$ minutes versus $30$ minutes. Moreover, the models prior work investigated are static and are learned a priori. Our work develops a new kind of a predictive model that is targeted toward representing physical skills and capability required for walking. The model can be revised online and gradually adapts to each specific trainee.

\section{Background}
We begin by describing how expert exercise coaches and physical therapists prescribe aerobic exercises. This information is useful for developing good knowledge representation for AI reasoning and adaptation. We then introduce our smartphone application - \textit{NutriWalking}. It is through this application that our AI coach interacts with its human trainee. Finally, we briefly describe the AI architecture we developed for real-time reasoning and coaching.

\subsection{Health Coaching}
The primary goal of our AI coach is to motivate sedentary, overweight individuals to walk regularly. Walking is one of the simplest aerobic activities and is usually the primary exercise experts recommend for a healthy lifestyle. Walking requires limited resources and is easily adaptable to individuals with varying capabilities. Importantly, walking is safe because it is a low-impact, familiar activity which minimizes the risk of injuries. These characteristics make walking an ideal exercise domain to study AI coaching methods.

\subsubsection{Dosage of Aerobic Exercise}: To be most effective, it is necessary that the prescribed dose of exercises be tailored to an individual's current physical capability and desired health goals. In clinical practice, the \textsc{FITT-VP} principle of exercise prescription is used by experts to tailor physical activity goals. The principle suggest changing \textbf{F}requency ($f$ per week), \textbf{I}ntensity ($i$ e.g., light, moderate, vigorous), \textbf{T}ime ($t$ per session), and \textbf{T}ype of exercise (in our case walking) to adjust or \textbf{P}rogress exercise \textbf{V}olume ($v = i \times t \times f$) per week that correlates with energy expenditure \cite{american2013acsm}.

There are various guidelines and clinical recommendations to determine the exercise volume that should be prescribed. To improve overall cardiovascular health, the American Heart Association (AHA) recommends at least $150$ minutes of moderate intensity exercise or $75$ minutes of vigorous intensity exercise per week or a combination of the two for adults. Given our objective to promote exercise behaviors in sedentary, overweight individuals, we chose the AHA recommendation of  at least $150$ minutes of moderate intensity exercise (given that our target population is sedentary) as
the target exercise volume (i.e., AHA goal) to be achieved by the trainee through agent coaching.  Measures of dose, intensity and physical capability are central to coaching trainees in walking.

\subsubsection{Intensity Measure}: Exercise intensity is described based on energy demands of physical activity. It is measured through caloric expenditure, e.g., a MET (metabolic equivalent) that is defined as the amount of oxygen consumed while sitting at rest \cite{jette1990metabolic}. A MET value for any physical activity is the energy cost expressed as a multiple of the resting metabolic rate. The Compendium of Physical activities \cite{ainsworth20112011} standardizes the assignment of MET values for different physical activities. Our method adopts MET values to determine activity intensity. In clinical practice, percentage maximum heart rate ($\%$HRmax) and/or a subjective rating of perceived exertion (e.g., Borg's Rate of Perceived Exertion Scale i.e., RPE) are often used to enable people to self-monitor exercise intensity. Although measuring $\%$HRmax requires training or pulse sensors, RPE is simpler because it involves reporting a subjective rating of how hard an individual feels like their body is working during exercise. RPE is highly correlated with actual HR during physical activity \cite{borg1998borg}. An important use of these measures is to adapt activity intensity to an
individual. Given lack of an HR sensor and constraints of mobile deployment, we modified the Borg's RPE scale to go from 1-5. The original Borg's RPE scale goes from $0$ (\textit{no exertion at all}) to $20$ (\textit{very hard e.g., sprinting as fast as you possibly can}). Typically, moderate exercise corresponds to a rating of about $11-14$ on the original RPE scale. Correspondingly, moderate intensity exercise is about $3$ in our scale. To facilitate understanding of the modified RPE scale, we used descriptors from the \emph{talk test} \cite{persinger2004consistency} along with tiredness descriptions in our scale. Ratings on our scale reflected:
\begin{enumerate}
\item no exertion - \emph{``Not tired at all''}
\item very light exertion - \emph{``Little tired: breathing felt easy''}
\item moderate exertion - \emph{``Tired, but can still talk''}
\item challenging - \emph{``Really tired: felt out of breath''}
\item impossible - \emph{``So tired: had to stop''}
\end{enumerate}

\subsubsection{Assessing Physical Capability}: In clinical settings, aerobic capacity is often precisely measured through elaborate tests of volume of oxygen consumption (VO2), or functional evaluations, such as $6$-minute walk tests that are more relevant in deconditioned (low initial capability) individuals. Given constraints of our mobile deployment, these objective, clinical evaluations are difficult and costly to implement. Therefore, we adapted a self-report instrument --- International Physical Activity Questionnaire (IPAQ), which is a reliable instrument to monitor physical activity levels \cite{Booth:2003ui}. In our questionnaire, individuals report duration and frequency of physical activities in a typical week. These questions probe duration and frequency for low intensity ($i_q = 3$ METs, such as stretching), moderate intensity ($i_q = 5$ METs, such as fast walking), and vigorous activities ($i_q = 8$ METs, such as playing a sport). Given these measures, an individual's initial physical capability can be computed as in Section \ref{sec:assessment}.

Given these measures, the AHA goal maps to $750$ METs-minutes weekly exercise volume ($i = 5$ METs, $d = 30$ minutes, and $f = 5$) with the activity performed at an average RPE of $3$ on our scale.

\subsection{NutriWalking Application}
\begin{figure}
    \centering
    \includegraphics[width=0.9\textwidth]{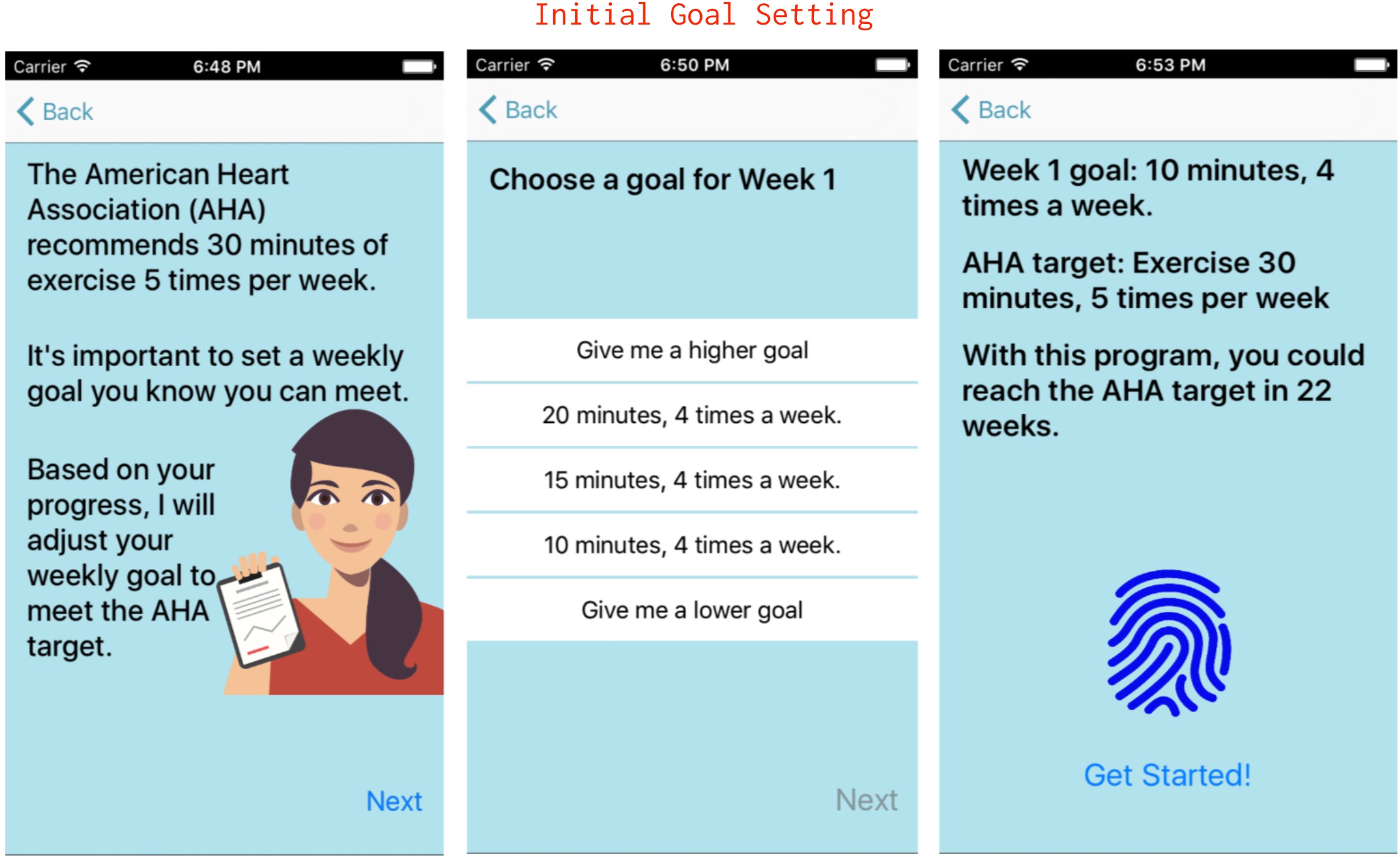}
    \includegraphics[width=0.9\textwidth]{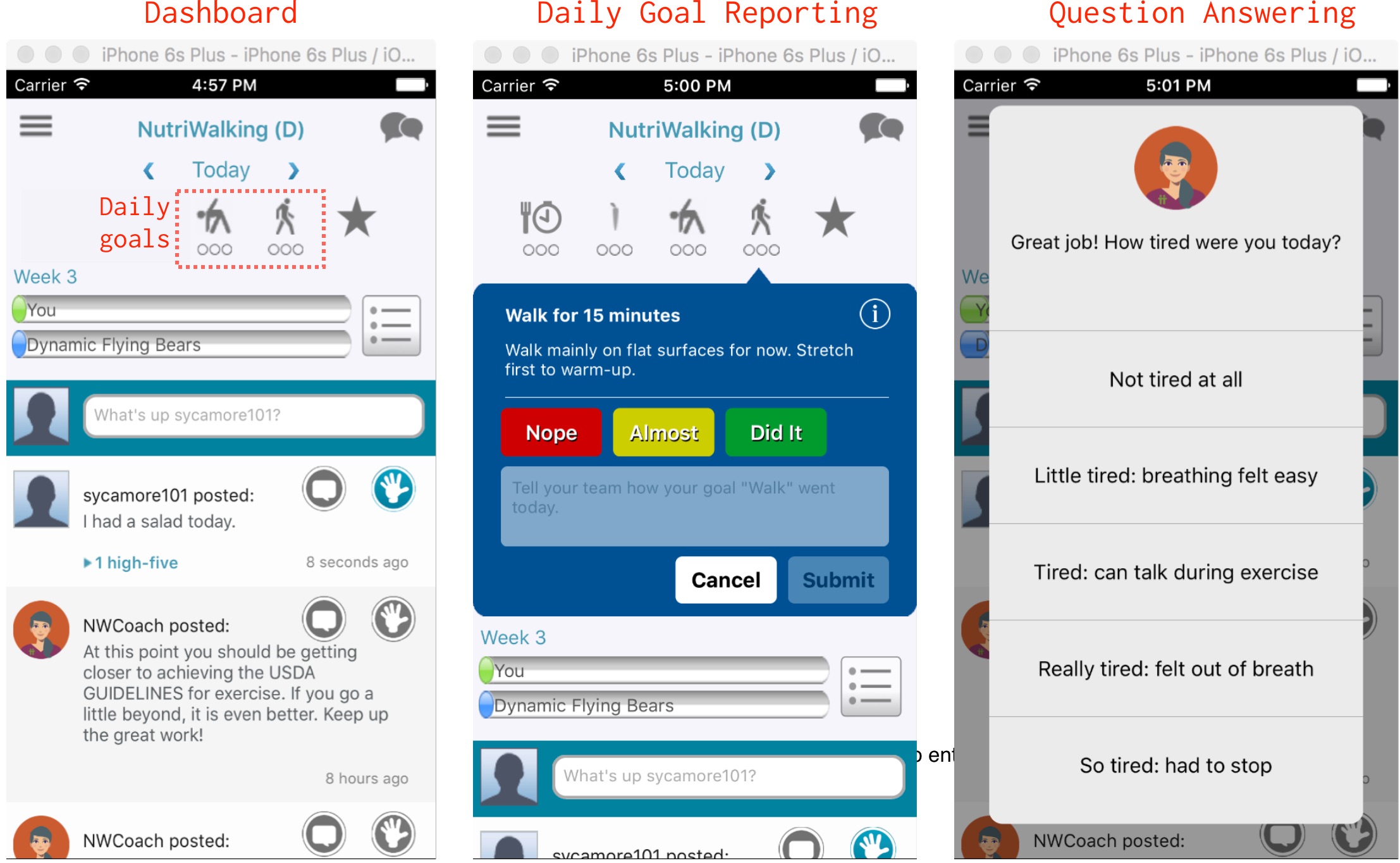}
    \caption{Various human-agent interactions \textsc{NutriWalking} - (\textit{top}) initial goal setting, (\textit{bottom left}) main dashboard, (\textit{bottom center}) daily goal reporting, (\textit{bottom right}) daily question answering}
    \label{fig:nutriwalkinga}
\end{figure}
\begin{figure}
    \centering
    \includegraphics[width=0.7\textwidth]{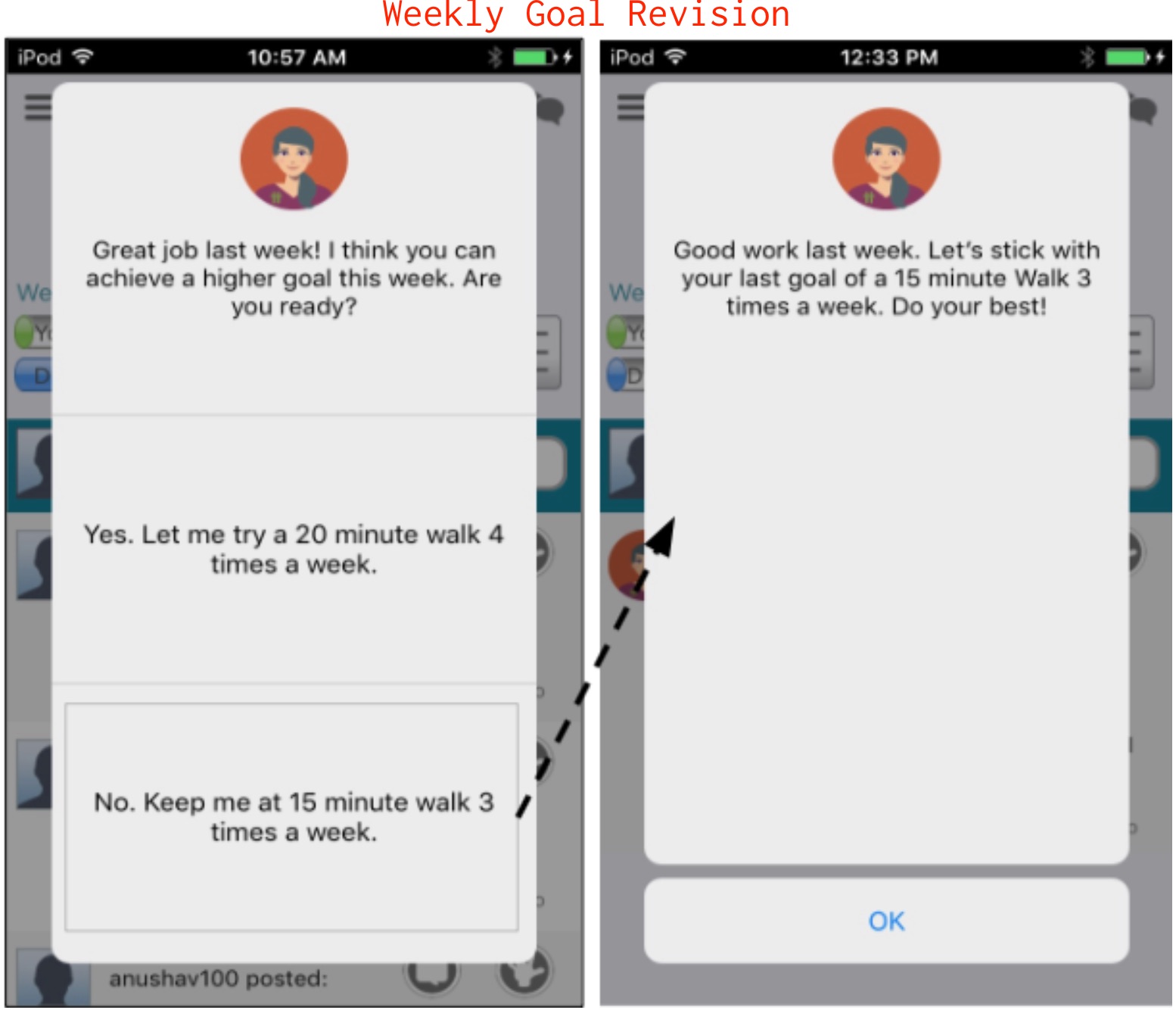}
    \caption{Various human-agent interactions \textsc{NutriWalking} - weekly goal revision}
    \label{fig:nutriwalkingb}
\end{figure}
To facilitate long-term coaching interactions with human trainees, we developed a smartphone application called  \textsc{NutriWalking} (Figures \ref{fig:nutriwalkinga} \& \ref{fig:nutriwalkingb}) that can be deployed to iOS users. The coach schedules personalized, relevant goals and monitors goal performance through various interactions in the application. The coach begins by informing the trainee about the AHA recommendation for weekly exercise and recommending a range of goals to choose from (Figure \ref{fig:nutriwalkinga}, top). This interaction ends when the coach affirms the goal and projects how long the trainee would need to reach the AHA target. Upon setting an initial weekly goal, the trainee has access to their dashboard (Figure \ref{fig:nutriwalkinga}, bottom left). The dashboard is the trainee's home-screen that contains information about their exercise goals as well as a social feed from their network (not a component reported on in this paper.) On this dashboard the trainee can view their daily activity goals (Figure \ref{fig:nutriwalkinga}, bottom left) and report on those goals by clicking on the appropriate answer (i.e., ``nope'', ``almost'', or ``did it'') (Figure \ref{fig:nutriwalkinga}, bottom center). This report triggers questioning (Figure \ref{fig:nutriwalkinga}, bottom right) through which the intelligent coach gathers information about the trainee's goal performance to adapt future goals. The dashboard also shows progress toward weekly goals, thereby allowing the trainee to track progress toward their long-term goal. At the beginning of every week, the coach uses the information collected during the previous week to evaluate whether or not the weekly goal should be revised to improve goal compliance. This goal is presented to the trainee to establish a joint goal between the trainee and the coach (Figure \ref{fig:nutriwalkingb}). \emph{NutriWalking} was not integrated with sensors that could be used for objective measurements of a person's physical activity and exertion, and relied primarily on self-reports; software development required for that integration was out of scope for this research at the time.

\subsection{AI Architecture}

\begin{figure}[b]
    \centering
    \includegraphics[width=1\textwidth]{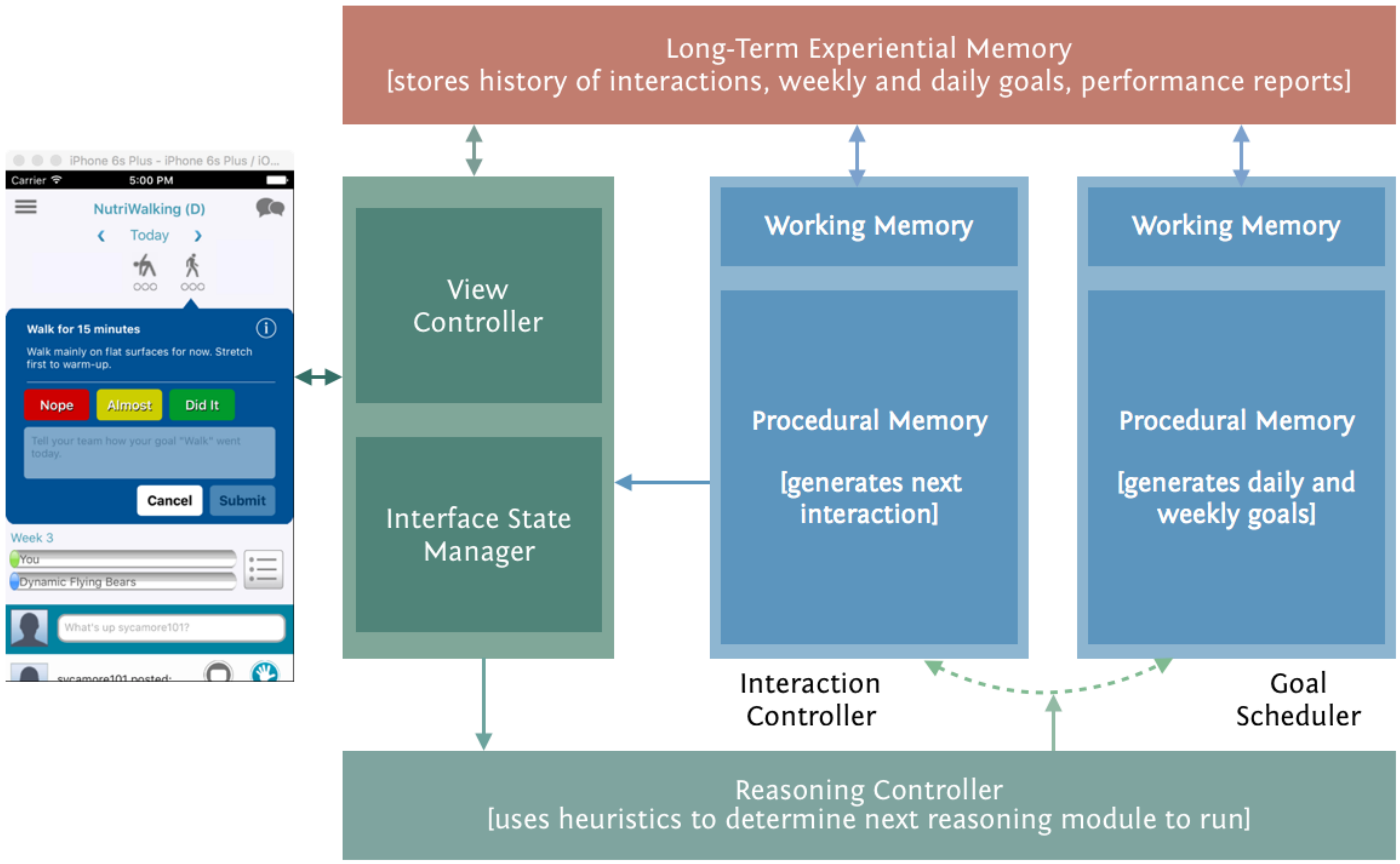}
    \caption{ \label{fig:architecture}A conceptual diagram of the AI architecture.}
\end{figure}

The intelligent coach is designed in a fashion similar to a symbolic, relational cognitive system \cite{langley2009cognitive}. Figure \ref{fig:architecture} shows the architecture of the intelligent coach. The coach employs the standard iOS view controllers and interface state managers to manage various application screens (Figure 2, left). The information acquired through interactions is stored in a database described below.

The coach has a specialized long-term memory (shown in red at the top in Figure \ref{fig:architecture}) that stores information from past interactions with the trainee as well as their performance on exercises recommended by the coach previously. This memory is developed using a relational database -  POSTGRESQL \cite{stonebraker1986} as well as objective-C code that stores information acquired through interactions and queries the database to retrieve useful information. The information stored in the memory is made available in the working memory of the two reasoners described below and drives their reasoning behavior.

The coach has rule-based reasoning modules for two primary behaviors - interaction and goal setting that rely on information from each other for reasoning shown in Figure \ref{fig:architecture} in blue (right two boxes in center row). These modules are developed using the CLIPS \cite{giarratano2005expert} rule-based engine. Each of these modules has a short-term working memory that is encoded as a graph. It contains the coach's current state which includes its hypotheses about the trainee's aerobic capability, its past experiences with the trainee, as well as the current state of interaction. Hand-written rules drive the coach's reasoning. Whenever a rule's left-hand side condition matches the working memory, it fires and its actions change the working memory. If several rules match at the same time, numeric preferences are applied to force an order. Interactive behavior is employed to assess a trainee's state, to gather more information about their performance, or to achieve joint agreement about goals. Interactions between the coach and the trainee are event-driven and may be initiated either by the trainee or the coach. Goal setting behavior looks at past interaction and performance data in the long-term memory and creates a schedule of walking goals. These goals are stored in the long-term memory for archival and future reasoning purposes.

The coach also contains a reasoning controller (Figure 2, bottom) which is implemented using objective-C. Its primary job is to determine which reasoning module the coach should employ - if it should run the interaction controller or the goal scheduler. It uses the current state in interface manager as well as some pre-encoded heuristics to determine which reasoning module to run.

\section{Computational Formulation of Goal Setting}
\label{sec:compu_form}
Having a goal is a crucial cognitive determinant of human behavior and performance \cite{locke2002building}. Success and failure to achieve a goal influences appraisals of other similar goals and motivation to pursue them. Not surprisingly, setting behavioral goals such as \emph{walk a mile in the evening} is one of the most promising strategies (or \emph{interventions}) employed in health behavior coaching \cite{shilts2004goal}. To be most effective, goals should be:
\begin{itemize}
\item \emph{difficult} yet \emph{attainable}: Goals must induce effort to improve performance but should not be too difficult to be successful at. For example, if an individual is consuming two servings of dairy products daily, a difficult yet attainable goal could be eating four servings of dairy products. For aerobic exercises in our case, difficulty can be expressed as the volume prescribed given the trainee's physical state. The attainability can be considered as how safe the prescribed goal is for a given physical state and how likely is it to be successfully completed.

\item \emph{specific}: Goals must provide a clear and narrow target for which the required type and amount effort can be estimated. General goals do not provide the basis for estimating or regulating effort and thereby are not pursued consistently. For example, a specific goal is to \emph{walk 1 mile in the evening after work} compared to a general goal to \emph{exercise more often}.

\item \emph{proximal}: Goals must be short-term, mobilizing effort in the near future. Long-term, distal goals make it easy to postpone effort. For example, \emph{walk at a brisk pace for 20 minutes tomorrow evening} is more effective than \emph{be more active this month}.
\end{itemize}

Following these suggestions, the coach is designed to maintain a fully specified schedule of walking exercises for seven contiguous days starting on the current day. A trainee can view this schedule in the NutriWalking smartphone application. On the days that the trainee is expected to exercise, the application shows the scheduled exercise (\emph{e.g., brisk walk}) and its duration (e.g. $20$ \emph{minutes}). Although our representation does not explicitly describe the quantities expressed in the goal setting theory, the proposed method and heuristics implicitly captures them.

The goal setting problem can be considered to be composed of two sub-problems: \emph{weekly scheduling} to determine the exercise volume to be pursued in a week given a long-term goal and \emph{daily scheduling} to distribute the weekly goal to specific days.

\subsection{Weekly scheduling}
Consider a trainee who has an aerobic capacity of $c_0$ at the beginning of the intervention and is advised to achieve the goal $g_n$ for a healthy lifestyle. The weekly scheduling problem is to generate a schedule of exercise goals $G_{[1, n]}= \{g_1, ... , g_n\}$ for weeks $[1, n]$. Following the FITT-VP principle, the weekly goal is represented as a tuple $g_w = (n,i,d,f)$ where $n$ is the exercise name, $i$ its intensity, $d$ the duration of a session, and $f$ sessions in a week. As the trainee achieves the exercise goal each week, their capability grows as a function of their prior capability and the exercise schedule $c_w = m(c_0, G_{[1, w-1]})$. A week's goal is $g_w$ selected such that it requires the capability $c_w$ for successful completion. This relationship is captured in a mapping function $g_w = r(c_w)$. Consequently, the goals get incrementally harder as the trainee's capability grows. At week $n$ the trainee can achieve the long-term goal $g_{n}$ which requires capability $c_n > c_0$.

\subsection{Daily scheduling}
Given goals for two consecutive weeks $w$ and $w+1$ and number of sessions completed in $w$, the daily scheduling problem at day $d_w$ in week $w$ is to select days in the interval $[d_w,d_w+7]$ on which sessions will be scheduled where some days in  $[d_w,d_w+7]$ may be in week $w+1$. This should be done so that the opportunity for the trainee to achieve their weekly goals is maximized and adequate rest days are scheduled between sessions.

\section{Trainee Aerobic Capability Model}
\label{sec:model}
Given weekly and daily scheduling, an optimal weekly schedule can be computed using a standard forward search algorithm. However, a few crucial challenges must be addressed. First, measuring a trainee's capacity $c_0$ through a mobile platform is neither straightforward nor precise. Experts rely on questions about a trainee's lifestyle as well as physical tests to estimate a trainee's capacity. To automate this process, several problems related to computer vision must be solved. Our coach uses a questionnaire about how active the trainee is in a typical week to generate an initial hypothesis about the trainee's aerobic capability. However, this assessment is error prone.

Second, as every trainee is different, the model $m$ required for scheduling cannot be fully specified during design time. It has to be fit to every individual who the coach trains. Additionally, the coach's goals should be reasonable even at the beginning of the program. This is a requirement because if the goals are too hard a trainee might injure themselves or if goals are too easy then the trainee might not be motivated to continue. Finally, an ideal solution for daily scheduling requires knowing how much time the person has available on each day and distributing the sessions accordingly. However, without knowledge of a trainee's schedule, it is hard to determine which days are ideal.

These challenges motivate an adaptive, knowledge-rich approach to designing the model and scheduling goals. The adaptability of our method alleviates the errors in assessing a trainee and the lack of a trainee's schedule. Although the initial goals may not be ideal due to assessment errors, the model can be refined based on observations made during training. To develop the parameterized model described below, we analyzed how experts prescribe exercises. The model encapsulates the structure experts rely on to prescribe exercises. The parameters can be revised online to fit different trainees.

We assume that people differ in two important ways: in their aerobic capability at the beginning of the program and in how quickly their physical capability can grow. The coach represents a trainee's aerobic capability $c_w$ as the exercise volume they can achieve in a week $w$. Given the intensity $i$ METs, the duration of a session $d$, and the number of sessions in a week $f$, their weekly aerobic capability is computed as $c_w = i \times d \times f$. Not only is this measure of physical capability standardized across various aerobic activities, it provides a direct mapping ($g_w = r(c_w)$) between the activity goals and the capability required to achieve those goals.

\begin{figure}[t]
\centering
\includegraphics[width=1\textwidth]{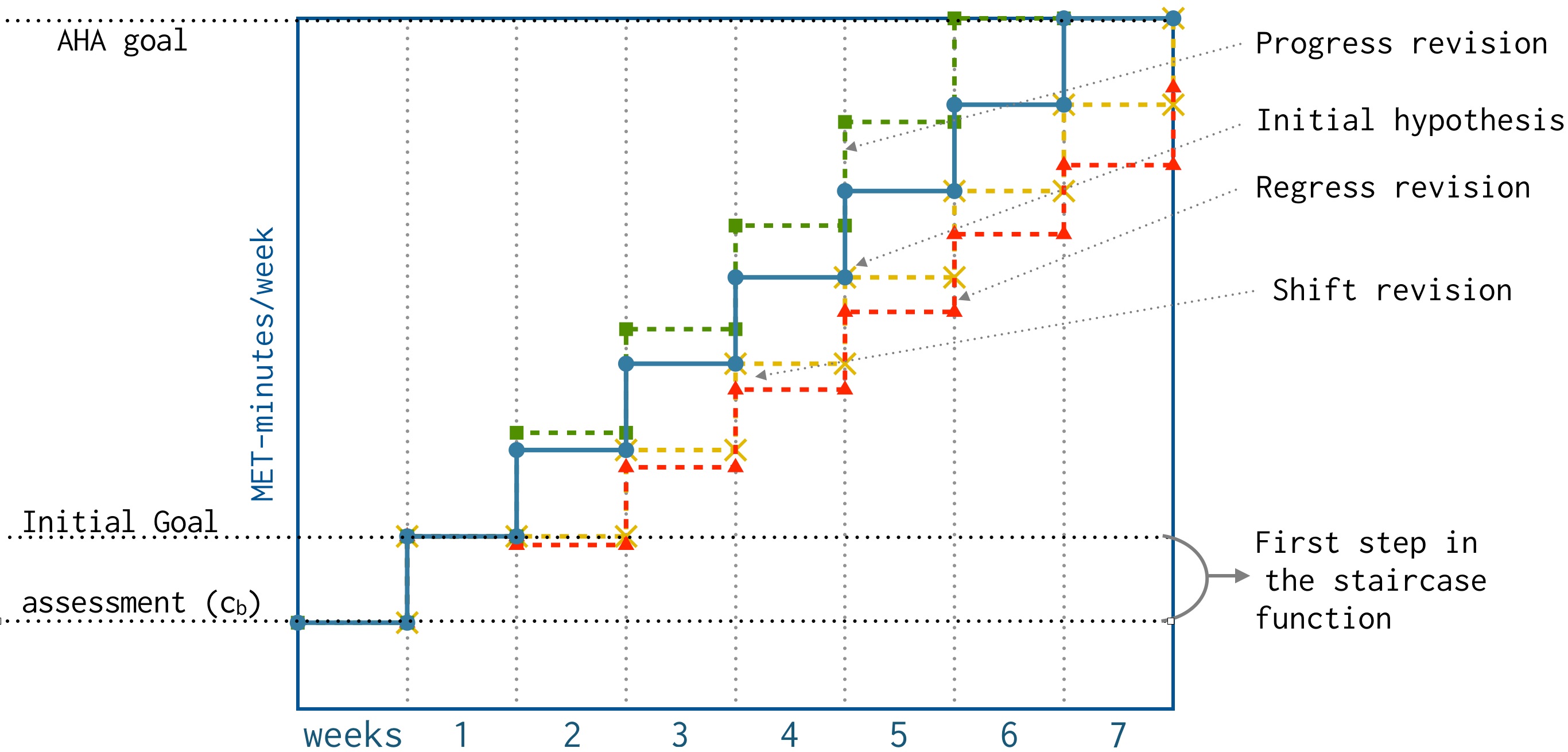}
\small
\caption{An adaptable model for aerobic capability. Blue line ($\CIRCLE$) represents the staircase model for growth in capability and green ($\blacksquare$), yellow ($\times$), and red ($\blacktriangle$) its revisions.}
\label{fig:staircase}
\vspace{-0.2cm}
\end{figure}

To capture weekly growth in aerobic capability, we employ a staircase function. The model assumes that a trainee's capability grows as a staircase function of equally spaced (1 week) steps of uniform height. Figure \ref{fig:staircase} shows some examples of such a model. The height of the function at week $w$ captures the capability $c_w$ in that week. The step height captures the coach's hypothesis about how quickly a trainee's capability can grow. The model can be
abstracted as the tuple $(c_0, c_n, s, o)$ where $c_0$ is the height of the floor of the staircase, $c_n$ the height of the highest step, $s$ the span, and $o$ is an offset. The model has two parameters that can be adapted by revising this tuple as follows:

\begin{enumerate}
\item \emph{change-step}: The step height can be revised by increasing (or decreasing) the staircase's span (on the x-axis) by $\delta$ weeks making it easier (or more difficult). For example, increasing the step size of the blue staircase function ($\CIRCLE$) in Figure \ref{fig:staircase} by decreasing the span $s$ by $1$ week results in the green staircase function ($\blacksquare$). This corresponds to a revision in the hypothesis about how quickly a trainee's capability can grow.

\item \emph{shift}: The staircase function can be shifted forward on the x-axis by $\delta$ weeks. For example, shifting the blue staircase function ($\CIRCLE$) in Figure \ref{fig:staircase} by $1$ week ($o = o + 1$) results in the yellow staircase function ($\times$). This corresponds to a revision in the next week's capability without revising the hypothesis about capability growth.
\end{enumerate}

The revisions can also be applied together. For example, the red staircase ($\blacktriangle$) is achieved by shifting the blue one ($\CIRCLE$) by a week and increasing its span by a week.

\section{Adaptive Goal Setting}
\label{sec:adaptive-goal-setting}
Here we describe how the AI coach employs interactions and goal setting to coach a trainee toward the AHA target.

\subsection{Assessment}
\label{sec:assessment}
Before planning a schedule of exercise goals, the AI coach must assess the trainee's baseline aerobic capability at the beginning of the program ($c_0$). This is critical for initializing the trainee model described earlier and consequently to schedule appropriate goals. Through a series of assessment questions, a trainee reports the duration and frequency of physical activity categories (with varying intensity) they undertake in a typical week. The AI coach aggregates these volumes into an assessment capability $c_0$ (Figure \ref{fig:staircase}). $c_0$ demarcates the floor (the interval before week $1$) of the staircase function. Volume corresponding to the AHA goal ($c_n$) demarcates the height.

The AI coach calculates a set of choices for the first week's goal by using the following values. For intensity, the AI coach uses the following walking exercises (in increasing order of difficulty):
\begin{enumerate}
    \item \emph{Moderate}: Walking at normal speed of $1$ mile in $30-45$ minutes. This has an intensity ($i$) of $3.0$ METs.
    \item \emph{Interval A}: $4$ minutes of \emph{moderate} and $1$ minute of \emph{brisk} walking for an average intensity ($i$) of $3.6$ METs.
    \item \emph{Interval B}: $2$  minutes of \emph{moderate} and $3$ minutes of \emph{brisk} walking for an average intensity ($i$) of $4.8$ METs.
    \item \emph{Brisk}: Walking while vigorously pumping arms at a speed of $1$ mile in $15-20$ minutes. This has an intensity ($i$) of $6$ METs.
\end{enumerate}
The duration ($d$) range considered in the study was $5-30$ minutes with $5$ minute increments and the frequency ($f$) ranged from $3-5$ days per week.

The coach selects the lowest intensity exercise ($i$) that can achieve the same volume as what is assessed. For this exercise, the coach determines the least duration ($d$) and corresponding frequency ($f$) will achieve the same volume as the assessed capability. These values $i$, $d$, $f$ map a trainee's capability to the exercises in the program. The goal choices for the first week ($g_1$) are computed by incrementally adding $5$ minutes to the duration until the maximum duration is reached. After compiling this set, the coach asks the trainee to pick a goal that they are most comfortable in attempting in the first week. The difference between $c_0$ and volume of the chosen goal determines the height of the first step, $h$ (step height in Figure \ref{fig:staircase}). Assuming uniform height steps, the span of the staircase is computed by $s = (c_n - c_0)/h$. This span is the projected time the trainee will take to reach their goal. People have a reliable estimate of how successful they can be at a task \cite{Bandura1994self}. Incorporating their choice in the model ensures that coach starts with a reasonable hypothesis that can be further refined.

\subsection{Daily Scheduling}
At the beginning of the week $w$, the coach distributes the $g_w.f$ sessions between week days such that the rest days (days with no scheduled walking) are uniformly spaced. The trainee is expected to report their performance every day. They can report that they successfully performed the activity (\emph{done}), that they tried the exercise but could not complete it (\emph{almost}), or that they did not do the exercise (\emph{nope}). For scheduling, \emph{almost} and \emph{nope} are equivalent because we want to provide ample opportunity to the trainee to achieve their goal. If successful, the original schedule is maintained. Otherwise, the coach redistributes $g_w.f$ in the remainder of week $w$ such that rest days are uniformly spaced. As the trainee moves through week $w$, the coaching starts scheduling week $w+1$ similarly to maintain a schedule for seven contiguous days. If the trainee is unable to achieve the daily goal, this rescheduling ensures that they get another chance to achieve it as long as there are enough days in the week.

When the trainee reports on an activity, the coach engages them in further interactions to gather more information about their performance. The trainee can report in three ways. On reporting \emph{done}, the coach asks the trainee to report how tired they felt during the exercise using the 5-point Likert RPE scale. The ideal exercise for the trainee should make them tired but they should still be able to talk ($3$ on the Likert scale). On reporting \emph{nope} or \emph{almost}, the coach asks the trainee to pick the reason why they did not do the activity by choosing an option from multiple-choice question. The available options include if they forgot about it, didn't have time, don't enjoy it, don't find it useful, and found the activity too hard. These interactions elicit information that is useful for the coach to revise the hypothesis if needed. The daily scheduling that occurs during the week can be considered an evidence-collection phase in which the coach observes the trainee's performance on a prescribed goal. This evidence is incorporated in the coach's reasoning.

Understanding why the trainee didn't comply with the scheduled exercise is important for providing coaching appropriate for the trainee. There are several reasons why a trainee may not complete a scheduled activity. It could be that they thought the exercise was too hard for them to pursue. However, there may be other reasons as well - the trainee lacks sufficient motivation, doesn't expect any internal or external reward, or simply forgot about it until it was too late. A good human coach adapts their coaching strategy based on determining why their trainee didn't complete the scheduled exercise. For example, if the exercise is too hard, it should be simplified to ensure success. Yet, if the trainee lacks motivation, a reminder of why they committed to exercise may be useful to improve compliance. The difficulty of the exercise should be revised only if non-compliance results from it being too hard. In other cases, different adjustments should be made. The current design of the intelligent coach is limited in such diagnostic reasoning required for effective coaching. We make a simplifying assumption that only if the trainee reports that the exercise was too hard, adaptions are made to their exercise schedule by the coach. Future extensions will enable the intelligent coach to reason about the full complexity of trainee behavior.

\subsection{Weekly Scheduling}
Based on the observations and interactions in week $w$, the coach may update the staircase model for weeks $> w$. This update is triggered by observations that deviate from expected performance of the trainee on the selected goal $g_w$. If in week $w$, the determined activity goal $g_w$ is appropriate for a trainee's capability $c_w$, it is expected that the trainee can achieve it with an average exertion of $3$. Any diversion from this should trigger a revision. The coach first determines whether the current model under- or over-estimates the trainee's capability and then adapts it as follows.
\begin{itemize}
\item \emph{regress} revision: The coach makes a regress revision if,
  \begin{itemize}
  \item the trainee is unable to complete at least $50\%$ of $g_w$ or,
  \item the trainee completes $> 50\%$ of $g_w$ but reports an average exertion $>= 4$ or there exist at least one report in which the reason for not completing an exercise is that it is \emph{too hard}
  		\end{itemize}
If these criteria are met, the coach assumes that the staircase model is not only overestimating the trainee's capability this week but it also overestimates how quickly the capability can grow. The coach revises the model by increasing the span of the staircase (a \emph{change-step} manipulation with $\delta = 1$) and thereby making the staircase less steep. This represents a revision in the coach's hypothesis about how quickly the trainee's capability can grow. The coach also shifts the staircase function by $\delta = 1$. As the trainee failed $g_w$, this revision ensures that the next week's goal will be easier than this week.

\item \emph{progress} revision: The progress revision occurs when the coach observes that the trainee completed at least $75\%$ of $g_w$ and reported an average exertion of $<=2$. This signals that the goal scheduled given the model's estimate of the trainee's capability and growth is too easy. Therefore, the model should be revised to reflect a faster growth in capability. The coach revises the model by decreasing the span of the staircase (a \emph{change-step} manipulation with $\delta = -1$) and thereby making the staircase harder.

\item \emph{shift} revision: The shift revision occurs when the trainee's non-performance is caused by something other than their aerobic capability. The coach makes this determination if the trainee completes $50\% - 75\%$ of $g_w$, the criterion for a regress revision was not met, and there exist at least one report in which the reason for not completing the daily goal is that the trainee was \emph{too busy}. This suggests that the $g_w$ may be appropriate and if given another opportunity, the trainee may be able to achieve it. The coach shifts the staircase function by $\delta = 1$ week to give the trainee another opportunity to complete the goal without revising the hypothesis about growth in their capability.

\end{itemize}
\noindent
\textbf{Weekly goal setting}:
Given a trainee's capability $c_w$ from the model, the coach computes the activity goal $g_w$ as follows. First, the coach generates possible combinations of intensity, duration, and frequency. For every activity of intensity $i$ under consideration and for every frequency value $f$, the coach computes the relevant duration value $d = c_w/(i
\times f)$ approximated to the closest multiple of five. Any combination in which the duration is higher than the maximum or lower than the minimum is rejected. If $w = 1$, the combination that matches the trainee's choice (in assessment) is set as the goal. For $w \neq 1$ the following filters are applied incrementally:
\begin{itemize}
\item if $c_w = c_{w-1}$, the combination matching the previous week's goal is selected as this week's goal.
\item if $c_w < c_{w-1}$ it implies that the previous week's goal was harder than what the trainee could achieve. Therefore, the combinations that are harder than the previous week's goal ($(g_w.i > g_{w-1}.i) \vee (g_w.i = g_{w-1}.i \wedge g_w.d > g_{w-1}.d) \vee (g_w.i = g_{w-1}.i \wedge g_w.d = g_{w-1}.d \wedge g_w.f > g_{w-1}.f)$) are rejected. From the remaining combinations, the easiest combination is selected to be this week's goal. Selecting the easiest goal ensures that a relatively safe goal is scheduled given the constraints derived from the model.
\item if $c_{w} > c_{w-1}$, it implies that the trainee can attempt a goal that is harder or at least equal to the previous week's goal. The combinations that are easier than the previous week's goal ($(g_w.i < g_{w-1}.i) \vee (g_w.i = g_{w-1}.i \wedge g_w.d < g_{w-1}.d) \vee (g_w.i = g_{w-1}.i \wedge g_w.d = g_{w-1}.d \wedge g_w.f < g_{w-1}.f)$) are rejected. As described earlier, the easiest combination is selected from the remaining combinations.
\end{itemize}

\section{Evaluation}
Evaluation of our interactive intelligent system presented a tremendous challenge. Not only does our system comprise a novel computational formulation of coaching, we also implemented specific human trainee-AI coach interactions in \textsc{NutriWalking} to facilitate coaching. Further, we expect coaching to be most impactful in a long timeframe ($4 - 8$ weeks). To ensure comprehensive evaluation of these various aspects, we pursued a four-pronged approach:
\begin{enumerate}
    \item First, in Section \ref{sec:adaptivity} \textbf{Adaptivity}, we study the characteristics of adaptation in our proposed algorithm via simulating behaviors of different types of trainees. This adaptivity evaluation ensured that our algorithms were reactive to variance in trainee behaviors and functioned appropriately.
    \item Second, in Section \ref{sec:suitability} \textbf{Suitability}, we evaluate the suitability of goals produced by our approach by studying their alignment with recommendations expert physical therapists make. This evaluation determined whether the design of our algorithms addressed the desiderata suggested by the goal setting theory -  proposing goals that are useful and safe.
    \item Third, in Section \ref{sec:usability} \textbf{Usability}, we study whether the \textsc{NutriWalking} application and the trainee-coach interactions incorporated in it were usable by our target population. This ensured that human trainees were able to interact with the application as expected and the information collected using these interactions was useful for personalizing exercises.
    \item  Finally, in Section \ref{sec:impact} \textbf{Observed Impact Over 6 Weeks}, we study the impact of our approach by deploying \textsc{NutriWalking} to a group of $21$ sedentary participants and observing its impact on their walking over $6$ weeks. This evaluation assessed whether our proposed method is able to coach  human participants in real time in their day to day lives.
\end{enumerate}

\subsection{Adaptivity}
\label{sec:adaptivity}
Before we deployed the coach within the NutriWalking application to the target trainee population, we examined how the implemented algorithm personalized the coaching plan to different trainees by adapting the exercise schedule.

\subsubsection{Method}
To assess adaptivity, we conducted a simulation study in which we simulated the behavior of three different kinds of trainees. We are interested in two different kinds of adaptivity: \emph{personal}, in which the coach adapts the exercise volume to each individual trainee and \emph{temporal}, in which the coach adapts the exercise volume as the trainee progresses in the coaching program. Our algorithm makes two primary assumptions about the variance across trainees - first, they may have varying levels of activity before their coaching program began and second, they may have varying growth per week (i.e. not every one advances at the exercise volume at the same rate). Given these variability dimensions, we simulated the following trainee profiles:
\begin{enumerate}
    \item \textsc{Trainee A}: This profile simulates a trainee who previously did not exercise regularly but is committed to their exercise program and follows the coach's recommendations. They have a moderate growth profile. Consequently, they achieve their weekly goal each week by performing the recommended exercise at an average exertion rate of $3$ on our exertion scale.
    \item \textsc{Trainee B}: This profile simulates a trainee who lacks prior experience in regular exercise. The trainee is not very committed to the exercise program and has a low growth rate. Consequently they are unable to meet their weekly exercise goal and report an exertion of $> 3$ when asked to exercise for longer than the prescribed duration of $> 20 $ minutes or at higher intensity than walking. In such conditions, Trainee B achieves only $50\%$ compliance in a week.
    \item \textsc{Trainee C}: This profile simulates a trainee who exercised regularly  before the coaching program but has a low growth rate (i.e., they walk for $10$ minutes, three times a week). Consequently, they are unable to meet their weekly exercise goal and report the exertion of $ > 3$ when asked to exercise for longer than the prescribed duration of $> 20 $ minutes or at higher intensity than walking). In such conditions, Trainee C also achieves only $50\%$ compliance in a week.
\end{enumerate}

The simulation was conducted as follows. The coaching program was initiated for a trainee and assessment questions were answered to indicate initial activity for the trainee. For example, \textsc{Trainee A} responses to all assessment questions by reporting $0$ minutes of activity. Then, daily reports were made for every day the exercise was scheduled based on the trainee profile. For example, all daily reports from \textsc{Trainee A}  were marked as \textit{"did it"} and the exertion scale was answered at level $3$ and they met all weekly goals recommended by the coach.

\subsubsection{Data Analysis}
For each trainee simulation, we collected the volume of exercise the coach recommended every week for $8$ weeks. A weekly recommendation has $3$ components: type of aerobic activity (and its MET value), duration of an activity session, and frequency in a week. These quantities are multiplied together to compute volume of exercise recommended in a week. \textsc{NutriWalking} stores all the goals recommended to trainees in a relational database on a server. This server was queried to extract recommended goal volumes for the 3 trainee profiles.

\subsubsection{Results}
\begin{figure}[h]
\centering
    \includegraphics[width=0.7\textwidth]{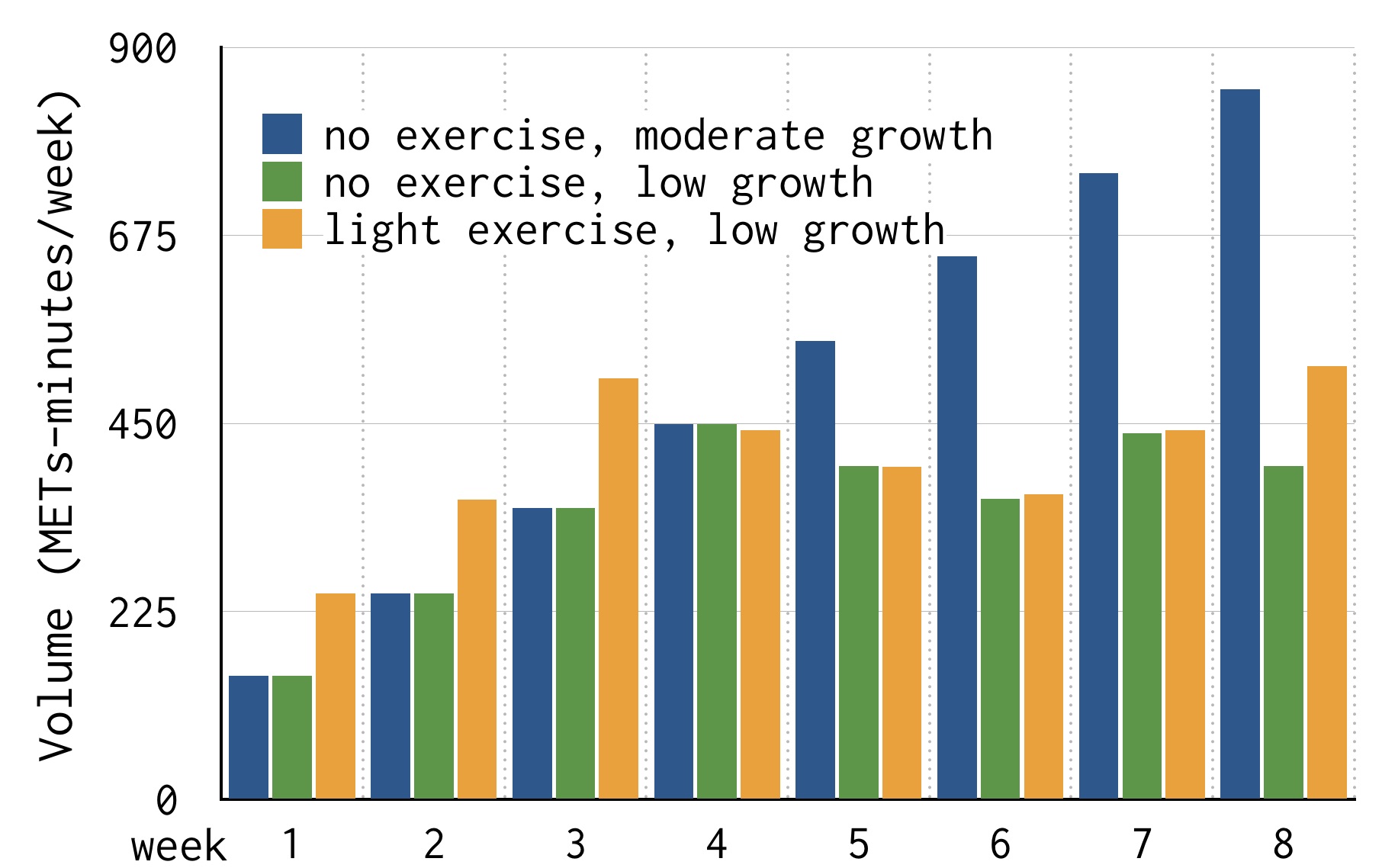}
  \caption{\small Prescribed exercise volume for $8$ weeks for $3$ profiles.}
  \label{fig:adaptive-result}
\end{figure}

The coach's adaptivity is demonstrated in Figure \ref{fig:adaptive-result}, which shows the goals scheduled by the coach for the three different trainee profiles over $8$ weeks. The dark blue bars represent the goals for a \textsc{Trainee A}. We see that the goals became incrementally harder (increase in volume) as time progresses. This trajectory can be compared to \textsc{Trainee B} (green bars) who did not previously exercise and whose aerobic capability grows slowly. By week $4$ the goal increases to $25$ minutes of moderate walk $5$ times a week. However, this trainee is unable to achieve this goal successfully. We can see that until week $4$, both trainees are assigned similar goals. However, week $5$ onward Trainee B is given easier goals as the coach revises its hypotheses about the trainee at week $5$ given observed performance until week $4$. \textsc{Trainee C} (light yellow bars) began at higher goals than Trainee A or B. But, their capability grows similarly to  Trainee B. Therefore, the coach makes the goals easier at week $5$. These results show that the coach can adapt the goals to changes in a trainee's capability (temporal adaptation) as well as to different trainees (personal adaptation).

\subsection{Suitability}\label{sec:suitability}
Next, we evaluated whether the coach produces suitable goals for the trainee. To define what \textit{suitability} means, we refer back to goal setting theory, which posits that to be maximally effective, coaching goals should be difficult yet attainable, proximal, and specific \cite{shilts2004goal}. Our computational formulation ensures that the goals are proximal and specific - the coach maintains a fully specified schedule of exercises for the next week. We conducted a study with $6$ domain experts, who are trained physical therapists, to evaluate whether the intelligent coach-selected goals are difficult yet attainable.
\subsubsection{Method}
The experiment was conducted by simulating various trainee profiles and observing the behavior of the coach. Given our domain analysis that trainees differ along two dimensions, we simulated $6$ profiles with $3$ levels of exercise activity in the beginning --- no, low, and moderate and $2$ levels of capability growth profiles --- low and moderate. We did not simulate high growth profiles as this is not characteristic of our target population that comprises of sedentary, overweight trainees. We asked $6$ physical therapists to judge the coach's performance. Each trainee profile was described to experts in terms of initial assessment, desired long-term AHA goal, history of weekly adherence, and average exertion scores. Experts were asked to judge coach-selected goals relative to $2$ control goals picked by a different expert such that they are reasonable and plausible given past history of trainee performance. For each week of the $8$-week program for all $6$ profiles, experts rated all $3$ goals on $6$-point Likert scales of \textit{safety}, \textit{usefulness}, \textit{likely to be completed}, and \textit{difficulty}. The neutral option on the Likert scale was deliberately omitted to force judgment in either direction. Importantly, the experts were blinded to the fact that these goal recommendations originated from an algorithm to reduce/avoid any unintentional biases. The experts rated these goals under the knowledge that they were prescribed by other experts such as themselves.

\subsubsection{Data Analysis}
In order to evaluate the differences in the \emph{correctness} of the AI coach goals as against the control ones, we first converted the expert ratings to a contingency table. For this, the $6$-point Likert scale levels were collapsed to obtain a binary classification (e.g., agree versus disagree for safety, likely to be completed, and usefulness while goal difficulty rated as difficult versus easy).  Further, the $2$ control goals were merged and compared to the coach-selected goals. To test whether the frequency distributions of agree versus disagree ratings were significantly different across conditions, we performed $\chi^2$-squared tests.

\subsubsection{Results}
\begin{figure}[h]
    \includegraphics[width=1\textwidth]{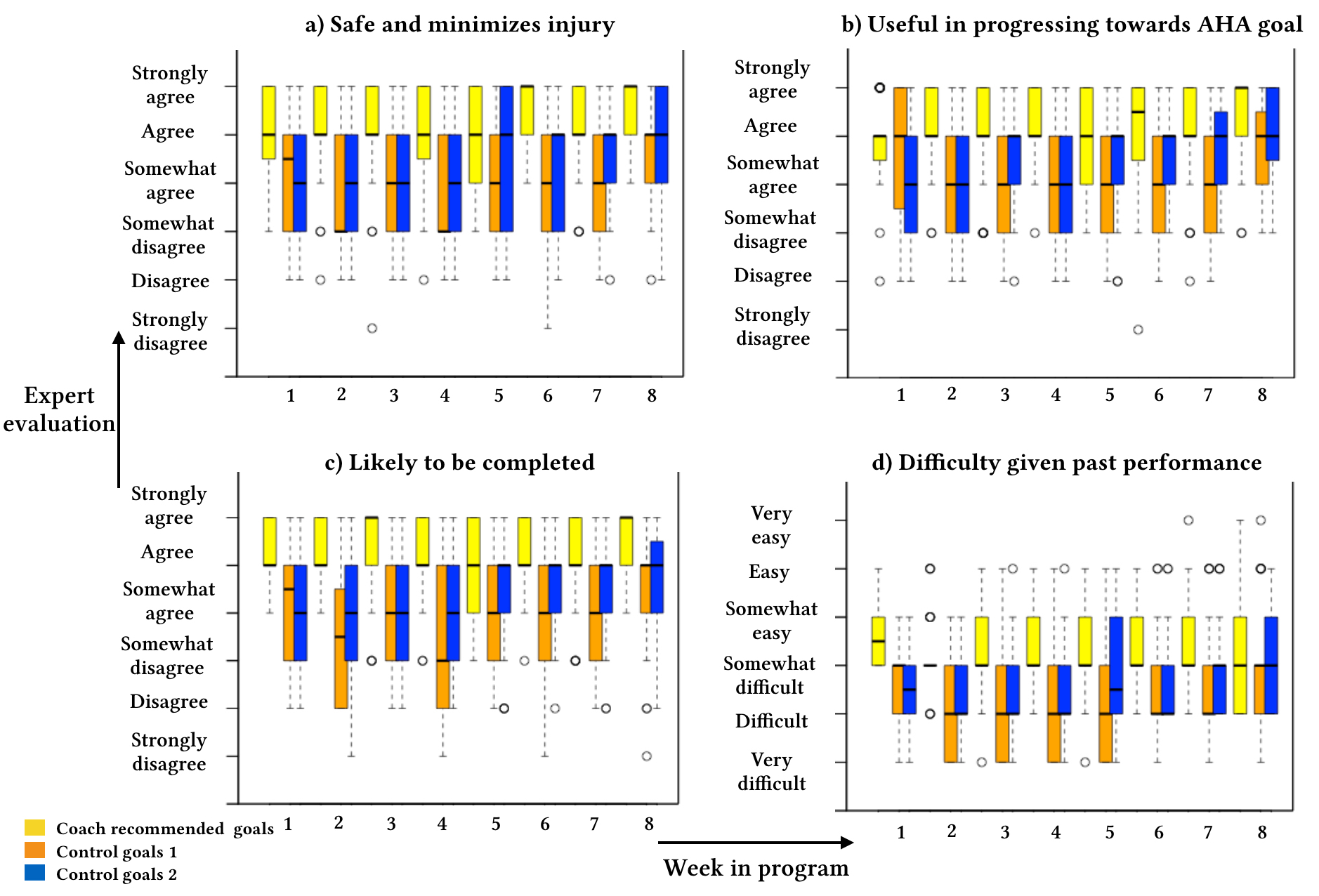}
  \caption{\small Box plots show expert ratings of weekly goals for an 8 week program. Boxes depict inter-quartile range, horizontal bar medians, dashed lines range, and circled outliers.}
  \label{fig:result}
\end{figure}

As shown in Figure \ref{fig:result}, experts rated the AI coach-selected goal higher on safety, attainability (i.e., likely to be completed), and usefulness compared to control goals ($1 \& 2$) for all $6$ profiles across $8$ weeks. These goals were also rated easier than controls. As shown in Table \ref{tab:results}, $\chi^2$-squared tests revealed that these differences in the ratings seen for coach versus control goals were significantly different ($p<0.001$).

Our results suggest that the coach selects goals that have a higher likelihood of being safer, are more likely to be completed, and useful to achieve a long-term goal compared with other plausible goals. Further, our method selects goals of appropriate difficulty given history of performance (see box plot $d$ in Figure \ref{fig:result}). It shows that the goals preferred by the coach are more likely to fall between \emph{somewhat easy} and \emph{somewhat difficult}. This demonstrates that the goals selected by the coach are reasonable for various trainee profiles and have a higher chance of being successfully achieved than other comparable goals.

\begin{table}[h]
    \renewcommand{\arraystretch}{1.2}
    \centering
   \begin{tabular}{|c|c|c||c|c|}
      \hline
      &\multicolumn{2}{|c||}{\textbf{Safe}}&\multicolumn{2}{|c|}{\textbf{Useful}}\\
      &\multicolumn{2}{|c||}{\tiny$\chi^{2}(1)=76.56$, \tiny$p<2.2e^{-16}$}&\multicolumn{2}{|c|}{\tiny$\chi^{2}(1)=40.21$,\tiny$p=2.267e^{-10}$}\\
      \cline{2-5}
      \small
      & \textbf{Agree} & \textbf{Disagree} & \textbf{Agree} & \textbf{Disagree}\\
      \hline
      \hline
      \textbf{Coach}& 266 & 19 & 266 & 22\\
      \hline
      \textbf{Controls}& 379 & 197 & 425 & 151\\
\hline
    &\multicolumn{2}{|c||}{\textbf{Likely}}&\multicolumn{2}{|c|}{\textbf{Difficult}} \\
      &\multicolumn{2}{|c||}{\tiny$\chi^{2}(1)=79.78$,\tiny$p<2.2e^{-16}$}&\multicolumn{2}{|c|}{\tiny$\chi^2(1)=23.32$,\tiny$p=1.37e^{-06}$} \\
      \cline{2-5}
      \small
      & \textbf{Agree} & \textbf{Disagree} & \textbf{Difficult} & \textbf{Easy}\\
      \hline
      \hline
      \textbf{Coach}& 275 & 13 & 156& 132\\
      \hline
      \textbf{Controls}& 393 & 183 & 409 & 167\\
        \hline
      \end{tabular}
     \caption{\label{tab:results}Contingency tables for binary expert ratings of coach-selected v/s control goals.}
\end{table}

\newpage

\subsection{Usability}\label{sec:usability}
In prior work \cite{Hartzler2016}, we assessed the usability of NutriWalking  in $15$ patients with a chronic health condition (Type 2 Diabetes Mellitus or depression). In a one hour session, we asked participants to use \textsc{NutriWalking} to complete usability tasks, too review adaptive coaching features through a cognitive walkthrough, and to respond to an acceptability survey with the System Usability Scale \cite{brooke1996sus}. Our findings indicated that:
\begin{enumerate}
    \item Overall, \textsc{NutriWalking} demonstrated good usability. Testing uncovered  minor enhancements that were later implemented.
    \item Participants found utility in the adaptive coaching features. Some reported that interacting with the coach on adaptive goal setting could provide them with \textit{"control"} to \textit{"help you take responsibility"}, with \textit{"more choice"}, and \textit{"allow you to set goals you know you can strive for"}. One participant expressed a concern about setting goals
too high and risk injury, while another thought that people might not challenge themselves enough to progress. There were concerns about responding to daily reporting questions being repetitive and taxing however participants thought that the process was easy to use.
    \item Participants reported good acceptability of \textsc{NutriWalking}, with a mean System Usability Scale score of $75/100$ ($70$ percentile rank).
\end{enumerate}
Through this study, we concluded that \textsc{NutriWalking} application was designed to be usable by our target population and could be deployed for longer-term study.

\subsection{Observed Impact Over 6 Weeks}\label{sec:impact}
Evaluation efforts described in previous sections established that the algorithm implemented has desirable coaching behavior and that  \textsc{NutriWalking} is usable and acceptable by our target population. However, these evaluations were conducted in highly controlled experimental settings with various simulations. An ideal coaching agent should integrate seamlessly in its trainee's daily lifestyle and provide interventions in the ecological context. To study if our proposed approach can meet this expectation, we conducted a $6$-week observational study with a sample of trainees from our target population. We assessed  how trainees interact with the coach and how the variation in this interaction impacts the performance of the coach and exercise compliance in trainees.

\subsubsection{Method}
This study was conducted collaboratively with our clinical research partners at Kaiser Permanente Washington. Pre-diabetic and type 2 diabetic patients with co-morbid depression were recruited through mailed study informational packets. A total of $21$ qualified participants (mean age $51 \pm 8$ years) volunteered to participate in the study. These participants completed a pre-intervention visit  when they were given instructions on how to install and use \textsc{NutriWalking} on their smartphones. Participants were also provided with a Fitbit physical activity tracker as an incentive for their participation. They used \textsc{NutriWalking} for the next $6$-weeks to interact with the coach and log their daily activity goals. The data provided to the application was backed up on a server maintained by the research team. This data was used to adapt weekly and daily goals by the AI coach as well as for the analyses described below. At the end of the study, participants completed a post-intervention visit and participated in an exit interview to provide subjective feedback about their experience using the app.

\subsubsection{Data Analysis}
As described earlier, we collected a variety of data about the trainee's behavior through trainee-coach interactions in the \textsc{NutriWalking} application. Particularly relevant are:
\begin{itemize}
    \item Assessment: This is a measure of the weekly baseline physical activity a participant was engaged in before they were enrolled in the study. Assessment was measured as a self-report by asking the participant questions about the  kinds of physical activity (\emph{intensity}), \emph{duration}, and \emph{frequency} they do in a week. An assessment activity volume was computed by multiplying these quantities.
    \item Proposed weekly goals: For each week of the study, the coach used data reported during the previous week (using methods proposed in Sections \ref{sec:compu_form}, \ref{sec:model}, \& \ref{sec:adaptive-goal-setting}) to determine the weekly goal volume for the trainee to strive for. The proposed weekly goal is represented as a tuple \emph{(activity type, frequency, duration)} and a \emph{volume} computed from the tuple.
    \item Weekly negotiation: At the beginning of each week, the coach proposed the goal computed to the trainee. It is presented as an activity (such as \textsc{Brisk walk} with a description), its \emph{frequency}, and \emph{duration}. The proposal can be \emph{increasing} the goal volume in comparison to last week's goal, \emph{decreasing} it, or \emph{stay}ing at the same volume. The participant can \emph{agree} or \emph{disagree} when there is a proposed change.
    \item Daily reports: Once the participant committed to the coach's proposal, activities are scheduled for the days in the week. The participant can look at their schedule and report physical activity. When they report having performed the scheduled activity, they are asked to provide a measurement of exertion on our RPE scale. Additionally, we measured self-efficacy and affective attitude. Self-efficacy is an individual's estimate of the likelihood of success at a goal. Affective attitude toward a behavior can be interpreted as how much a trainee likes or dislikes performing it. Both of these constructs have been found to be relevant in the context of behavior change \cite{tobias2009changing}. Here, we collected these data for subsequent analyses to understand how they relate to performance in the program.
    \item Weekly performed exercise volume: For each week, volume of exercise performed can be calculated by looking up the number of days successful reports were made and multiplying it by the intensity and duration of exercise.
\end{itemize}

Using these data, we conducted the following analyses to evaluate the impact of our proposed approach:
\begin{itemize}
\item In section \ref{sec:qual-feedback} \emph{Observations of app use experience}, we summarize findings from the exit interviews in which  participants provided qualitative feedback about their experience using the NutriWalking application.

\item In section \ref{sec:goal-setting-demo} \emph{Observations of goal setting adaptations}, we use the proposed weekly goals, performed exercise volume, and RPE to demonstrate goal adaptation and exercise performance for two trainees.

\item In section \ref{sec:changes-report-behavior} \emph{Changes in reporting behavior}, we did a preliminary investigation of participant survival in terms of their app use (daily reporting) behaviors by summarizing the changes in frequency of reporting across the $6$-week intervention period.

\item In section \ref{sec:changes-ex-behavior} \emph{Changes in exercising behavior}, we analyze how proposed goal and performed exercise volumes change during the course of study using several mixed-effect linear models of the form $y = \alpha + \beta x + \gamma p + \epsilon$, where $y$ is the dependent variable, $x$ is the independent fixed-effect variable, and $p$ is a random effect variable corresponding to a participant. $\alpha$ is the intercept, $\beta$ and $\gamma$ are regression coefficients, and $\epsilon$ is the error term. Participants were included as random effects in these models to account for individual differences in mobile application use, interpretation of measurement scales, and acceptability of coaching. The main dependent variable, weekly exercise volume, was regressed on the following independent variables: week number in the study and weekly proposed goal volume.

\item In section \ref{sec:assessment-impact} \emph{Impact of assessment interactions}, we study the impact of self-reported assessment on exercise performance. To do this, we define \textit{optimism} as how much a trainee overestimates their exercise capability beyond their first week's performance.

\item In section \ref{sec:weekly-goal-impact} \emph{Impact of weekly goal negotiation}, we study the impact of weekly goal negotiation on exercise performance. To do this, we performed $\chi^2$ and Fisher's exact tests to study the difference in success and failures in daily reports under various negotiation conditions.

\item In section \ref{sec:daily-reporting-impact} \emph{Impact of daily reporting}, we study the impact of daily reporting on exercise performance. We fit a linear regression model of the form $y = \alpha + \beta x + \epsilon$. Here $y$ is the reported successes and $x$ is the matrix of independent variables measured in the daily reporting - rate of perceived exertion (RPE), self-efficacy, and affective attitude. $\alpha$ is the intercept, $\beta$ regression coefficient, $\epsilon$ is the error term.
\end{itemize}

\subsubsection{Observations of app use experience}
\label{sec:qual-feedback}
Overall, participants were able to use the app through the $6$-week intervention study without too much difficulty. Several participants engaged with the app regularly by reporting their exercise completion. Participants' feedback about the app suggested that the app provided an easy way to access a regular exercise routine ``\emph{without having to worry too much}". The structured daily exercise goals could have reduced the cognitive load for some participants to plan exercises and could consequently increase the likelihood of exercise compliance in those with low initial physical activity levels. Moreover, some indicated that using the app also made them feel more accountable towards completing their walking exercise.

\subsubsection{Observations of goal setting adaptation}
\label{sec:goal-setting-demo}

\begin{figure}[b]
    \centering
    \includegraphics[width=1\textwidth]{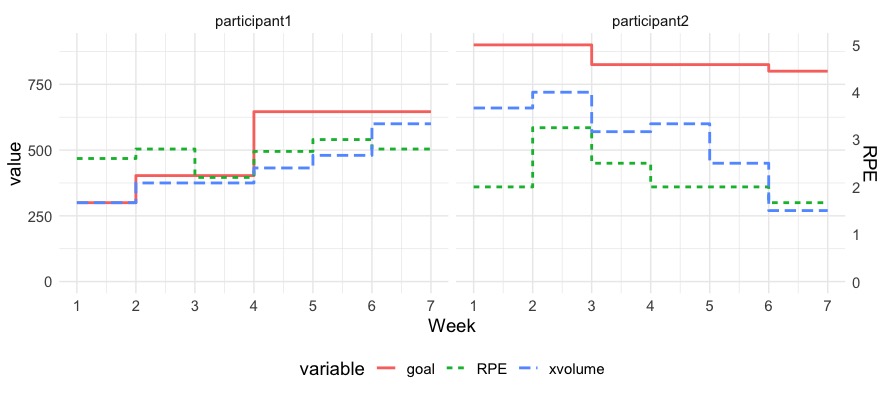}
    \caption{Weekly goals, exercise volume performed (xvolume), and rate of perceived exertion (RPE) of two participants for the length of the study.}
    \label{fig:behavior}
\end{figure}

We begin by showing how adaptive goal setting results in different goals and consequently different coaching experience for different trainees. Figure \ref{fig:behavior} shows the goal volumes (shown in red), performed exercise volumes (shown in blue), and reported rate of perceived exertion (shown in green) for two different participants. Participant $1$ began at a fairly low goal volume (about $300$ MET-mins/week, $20$ minute of moderate walking $5$ times a week). They were able to achieve this exercise volume in week $1$. The coach recommended a higher goal for week $2$, which they accepted and achieved. For week $3$, the coach recommended a yet higher goal. However, the participant chose to work on the goal from week $2$. As shown in figure 6, participant 1 was able to achieve that goal volume in week $3$. As they easily achieved these goal volumes, the coach made a progress revision for week $4$ and recommended increasing the intensity of walking to $25$ minutes of interval walk B (see section \ref{sec:assessment} for definition) which the participant agreed to. From then on, participant 1 was able to achieve at least $50\%$ of the goal volume and report RPE $< 4$ on an average. Consequently, in weeks $5$ and $6$, the coach recommended increasing the duration to $30$ minutes. However, the participant did not accept this recommendation and chose to keep the goal volume constant. During the coaching period, the participant increased their walking activity from $20$ minutes of regular moderate intensity walk to $25$ minutes of interval brisk walk B.

Participant $2$ began with the maximum goal volume $900$ ($30$ minutes of brisk walk $5$ times a week) that can be prescribed by the coach. The participant was unable to achieve even $50\%$ of this volume, but reported an average RPE of $<3$ in that week. The coach recommended reducing the duration to $25$ for week $2$, which the participant disagreed with and chose instead  to work on the original goal. In week $2$, the participant achieved a higher volume than week $1$ but also reported a higher RPE. Consequently, in week $3$ the coach recommended lowering the duration to $25$, which the participant agreed with. In week $3$, participant 2 reported an even lower RPE score. Judging that the exercise was too easy for them, the coach recommended increasing the duration. The participant disagreed with this recommendation. For the rest of the weeks, participant $2$ chose a goal of $25$ minutes of brisk walking $5$ times a week. However, their achieved exercise volume continued to drop along with reported RPE. We think that participant $2$ was either too optimistic in their goal setting and incorrectly reported their RPE as low, and our algorithm failed to account the optimism. The alternative explanation is that the prescribed intensity of exercise was too easy for this participant, so they were not adequately motivated to continue exercising. This limitation will need to be addressed in future work, perhaps with additional instructions before starting the program and augmenting the model with more quantitative activity metrics.

\subsubsection{Changes in reporting behavior}
\label{sec:changes-report-behavior}
\begin{figure*}[b]
   \includegraphics[width=.7\textwidth]{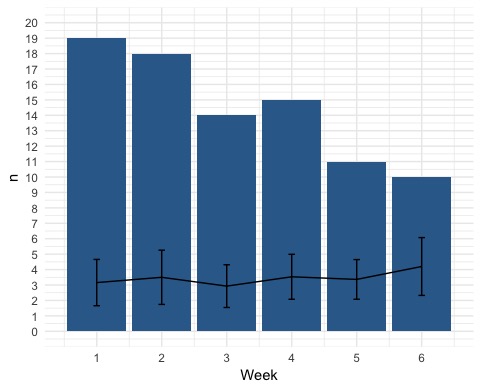}
    \caption{Number of trainees (n) reporting exercises in the application for each week in the study. The black line shows the mean and standard deviation of the number of reports per participant.}
    \label{fig:survival}
\end{figure*}

Figure \ref{fig:survival} shows the number of participants who reported their walking exercises in the NutriWalking application each week of the study. It also shows the mean and standard deviation of the number of reports made by each participant per week. While the number of active participants decreased every week, the number of reports active participants made every week did not change significantly (mean $3.44$, SD = $1.54$). This finding suggests that participants towards the end of the study were still regularly engaging with the coach.

\subsubsection{Changes in exercising behavior}
\label{sec:changes-ex-behavior}
The impact of coaching on participants' exercise behavior shown in Table \ref{tab:goalvolume}. As a sanity check, we regress goal volume on number of weeks in the study using a mixed-effects regression model. We see that goal volume (column $1$) for participants increased with the number of weeks in the coaching program. This is expected becasue the coach is designed to help people increase their exercise volume and it does so by helping people set higher goal volumes every week. In column $2$, we see that performed exercise volume increases with the number of weeks in the study as well. In our main analysis in column $3$, we see that the increase in exercise volume is explained by an increase in the goal volume every week. In fact, after controlling for the goal volume, the number of weeks spent in the study has a weak negative impact. Our regression coefficients indicate that increasing the goal volume by $100$ MET-mins/week for the next week increases the performed exercise volume by $61.8-0.487 = 60.313$.

Based on these findings, we can conclude that the coach works as expected. It gradually increases the goal volume for a participant every week. This increase in goal volume has a positive impact on the amount of exercise performed by the participant.
\begin{table}[t]
    \centering
     \def\arraystretch{1.1}%
    \begin{tabular}{cccc}
        \hline
        \hline
         &  (1) & (2) & (3)\\
         \textbf{Independent Variables}$\downarrow$ & Goal Volume & Performed Exercise & Performed Exercise\\
         \hline
         Week & 9.608*& 12.392* & -0.487* \\
         & (5.166) & (12.202) &(12.007)\\
         Goal Volume &&& 0.618*** \\
         &&&(0.119) \\
         \hline
         Mean Dependent Variable & 601.098 & 392.250 & 392.250 \\
         & (23.138) & (24.830) & (24.830) \\
         \hline
         Random effect & \checkmark & \checkmark & \checkmark \\
         Marginal $R^2$ & 0.004 & 0.005 & 0.378\\
         Conditional $R^2$ & 0.868 & 0.662 & 0.639  \\
         \hline
         \hline
    \end{tabular}
    \caption{Mixed-effect linear regression models for goal volume (column 1) and performed exercise volume (column 2). Volume is measured in MET-mins/week. The numbers in parentheses are standard errors. *** p $< 0.001$, ** p $< 0.05$, * p $< 0.1$}
    \label{tab:goalvolume}
\end{table}


\subsubsection{Impact of assessment interactions}
\label{sec:assessment-impact}
\begin{figure*}[h]
    \centering
    \subfloat[]{\includegraphics[width=.52\columnwidth] {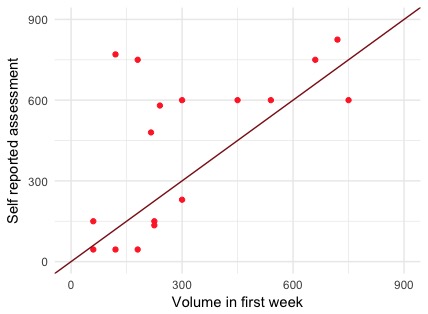}\label{fig:opt_rep}}
    \hfill
    \subfloat[]{\includegraphics[width=.46\columnwidth]{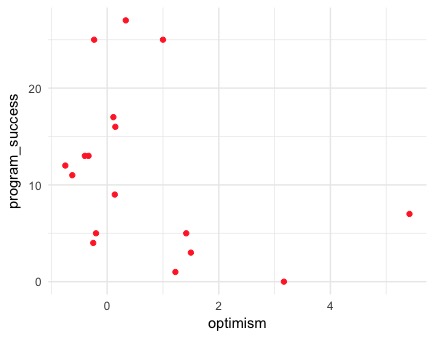}\label{fig:opt_success}}
    \caption{(a) Scatter plot of self-reported assessment vs first week exercise volume actually performed by all participants. The reference line is drawn at $45$ degrees. (b) Scatter plot of number of success reports vs optimism for all participants. Optimism is defined as (self-reported assessment - first week exercise volume)/first week exercise volume.}
\end{figure*}

We analyzed the impact of assessment interactions that occur when  participants begin using the NutriWalking application. Figure \ref{fig:opt_rep} shows the relationship between what participants report as their typical weekly exercise volume and how much exercise they performed in the first week of the study. Several participants were optimistic in their estimate (plotted above the $45$ degree reference line). This finding suggests that the self-report instrument used for assessment is error-prone. This is a concern because the coach depends on this measurement to be accurate for initializing the trainee model. Further, the adaptation technique does not let the cumulative height of the staircase model drop below the assessment level. This constraint becomes a critical issue when a participant is overly optimistic about their capability because all the goals recommended by the coach will be too difficult for the participant to succeed.

The problem caused by optimistic self-assessment is evident from Figure \ref{fig:opt_success}. We define \emph{optimism} as (self-reported assessment - first week volume)/first week volume. Figure 8b shows the relationship between the number of success reports a participant made during the coaching period and their optimism. Participants who are very optimistic are less likely to be successful at recommended exercises. To ensure robustness in this situation, future revisions made to the model must be based on more accurate information such as the performance in the first week.

\subsubsection{Impact of weekly goal negotiation}
\label{sec:weekly-goal-impact}
\begin{figure}[b]
    \centering
    \includegraphics[width=0.6\textwidth]{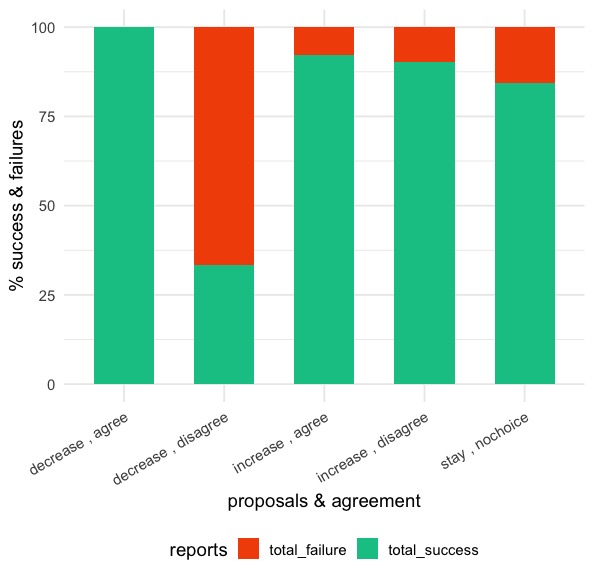}
    \caption{Proportion of success, failure, and absent reports made under varying levels of proposed revisions by the coach and agreement by the trainee. Pearson's $\chi^2 =  34.724$, df $= NA$, p-value $= 0.0004998$ with simulated p-value (based on $2000$ replicates). Fisher's exact test has a p-value $=2.223e-05$. Both tests suggest that the hypothesis that distribution of reports are independent of proposed revisions \& agreement can be rejected.}
    \label{fig:weekly_negotiation}
\end{figure}
We now report on the impact of the coach-trainee joint goal setting interactions that occur every week. Figure \ref{fig:weekly_negotiation} shows the distribution of success and failure reports under various conditions. When the coach proposed to decrease the exercise volume and the participants agreed, participants are largely successful. On the contrary, when the coach proposed to decrease the volume and the participants disagreed,  participants' success rate goes down significantly (Fisher's exact test has p-value $<0.005$). This finding suggests that the coach's recommendation to reduce the difficulty of exercise was appropriate.

The success rate when the coach proposes to increase exercise volume and participants agree is similar to when participants disagree and to when the coach proposes to stay at the current exercise volume. When the trainee disagrees with the coach's recommendation to increase the exercise, the participant selects the previous week's goal to pursue the current week. As the participant  was successful earlier, it is not surprising that they are successful again. Similar success rates in the agree condition suggest that the coach's recommendation to increase the exercise volume was not too aggressive.

\subsubsection{Impact of daily reporting}
\label{sec:daily-reporting-impact}
\begin{table}[b]
    \centering
    \def\arraystretch{1.1}%
    \begin{tabular}{cccc}
        \hline
        \hline
         &  1 & 2 & 3\\
         \textbf{independent variables}$\downarrow$ & success proportion & failure proportion & exercise volume\\
         \hline
         RPE & -0.103*& 0.080* & -23.51 \\
         & (0.061) & (0.046) &(47.28)\\
         Self-efficacy &-0.001 &-0.070 & -43.63 \\
         &(0.0714)&(0.053)&(54.92) \\
         Affective attitude & 0.028 &0.035 & 83.84** \\
         & (0.049) &(0.037) &(37.86)\\
         \hline
         Intercept & 0.614 & 0.141 & 197.04 \\
         & (0.340) & (0.253) & (261.88) \\
         \hline
         $R^2$&0.082&0.069&0.113\\
         \hline
         \hline
    \end{tabular}
    \caption{Linear regression models for proportion of success reports in a week (column $1$), proportion of failure and almost reports (column $2$), and performed exercise volume (column $3$). Proportions range from $0-1$. Volume is measured in MET-mins/week. The numbers in parentheses are standard errors. *** p $< 0.001$, ** p $< 0.05$, * p $< 0.1$}
    \label{tab:dailyreporting}
\end{table}
Table \ref{tab:dailyreporting} shows the impact of measured exercise performance variables: RPE, self-efficacy, and affective attitude on how successful a participant was and how much volume of exercise they performed in a week. The impact is measured through a linear regression. We see that RPE weakly influenced the proportion of successes and failures in a week. If the exercise assigned by the coach was too hard on the RPE scale, it was less likely to be successfully performed. This result suggests that using the RPE scale to regress the trainee model and revise exercises for the next week was a useful design choice. The volume of exercise performed during a week was significantly influenced by the affective attitude reported by the participant (t-test with p-value < $0.05$. Our hypothesis is that a participant who likes walking is likely to walk more and at a higher intensity, and therefore their exercise volume is higher. Self-efficacy did not have any significant impact on any dependent measure of successful behavior in our data.


\subsubsection{Summary of observed impact}
From our $6$-week long observational study of participants selected from the relevant population, we can infer the following:
\begin{enumerate}
    \item Participants were able to use \textsc{NutriWalking} without too much difficulty and many engaged with the application regularly. The coach demonstrated adaptive behavior with real participants (as opposed to simulated trainees).
    \item Several Participants increased their weekly exercise volume in response to the coach's goal recommendations.
    \item Interactive assessment may have caused participants to optimistically overestimate their capabilities leading them to fail more often. Future variations of the coach will focus on developing assessment interactions that reduce the potential for over-estimation.
    \item Analysis of weekly goal negotiations shows that our adaptive goal setting algorithm produces appropriate goals that are personalized to each trainee. Additionally, weekly goal negotiation provides a way for the participant to provide feedback to the coach and consequently control their exercise schedule while nudging them toward more exercise.
    \item Analysis of daily reporting suggests that RPE is a useful measurement to rely on for adapting the trainee's goals.
\end{enumerate}

\section{Limitations and Discussion}
This paper proposes a computational model to adapt aerobic exercise goals such that they are personalized to a trainee's capability and experience. The algorithms we developed are novel and are shown to be useful in our evaluations with a relevant population. However, the work presented here is not without limitations. Below, we summarize the limitations of the system and discuss how to address them in future research.

\subsection{Algorithm}
A primary limitation is that the model cannot be adapted such that the predicted capability in a week $w >= 1$ is lower than the initial capability measurement $c_0$. Therefore, if the initial capability is overestimated to a high degree, the coach may never recover from the error. Similarly, the staircase model is constrained to be adapted by at most a week, ($\delta = 1$) despite the method allowing for any arbitrary number, which precludes quicker adaptations. These are empirical questions that can be answered by deploying the coach in a larger sample in the future in which various values for these constants can be evaluated to optimize for maximal impact. While this is a limitation, it also highlights the strength of our approach - the computational models posits very specific hypotheses about a trainee's exercise capability which can be evaluated empirically.

Another limitation is reliance on self-assessment for benchmarking a trainee's capability. The instrument we used was associated with trainees' overestimating their capabilities and consequently disrupting the coach's adaptation. Ideally, assessment needs to be more accurate than the self-reported measurement used in the study. The coach can employ evaluation tests such as the \emph{1-mile walk test} \cite{weiglein20111} and use the data from wearable sensors to accurately estimate a trainee's current physical capability.

A limitation of the expert study is that experts were not probed about the rationale underlying their judgments. Insights so derived in the future could inform formulating a more expressive model of growth in aerobic capability as well as elicit heuristics that experts employ to adapt exercises for different trainees (personalized adaptation) as well as over time (temporal adaptation).

\subsection{Platform}
The current platform is designed for IOS and has been developed using Apple's mobile development suite. This limits deployment to only iPhone users who constitute a minority of smartphone users world-wide. However, the system proposed here can be adapted to other interactive platforms. An interactive AI application has two related but separable components - a core intelligent reasoning module and a reasoning module that deals with interaction with a human and is interfaced with the intelligent reasoning module. The intelligent reasoning module presented in this paper is general and can be deployed directly across platforms. The interactive module presented here, on the other hand, is specific to the smartphone embodiment and has to be adapted to the modality of interaction. For example, if the coach is deployed to a conversational system such as Alexa or Google Home, the interactions will need to be adapted to gather relevant information via dialog.

\subsection{Computational Model for Behavior Change}
Human behavior is  complex with several interacting determinants. To tease these determinants apart and develop interactive, adaptive methods to influence all of them is a tremendous challenge. This paper takes a small, principled step towards comprehensive behavior change systems and is in no way complete. Here, we focus only on aerobic exercise capability, a trainee's estimation of their capability (measured via self-efficacy), and how regular exercise impacts both of them. We show that a system designed with this consideration holds promise as a behavior-change tool. Health-behavior change literature identifies several interventions based on constructs such as self-efficacy \cite{Stacey2015}, implementation intentions \cite{belanger2013meta}, self-affirmations \cite{falk2015self}, and motivational interviewing \cite{miller2012motivational} that produce large effect sizes in positively influencing health behaviors. These theories impact other determinants of human behavior, such as memory, affect, and commitment. Our research aims to study these determinants, implement computational models, and design intelligent methods that influence them to develop a comprehensive system to aid behavior change. Other work \citep{pinder2018digital} also aims to develop integrated behavior change theory based on a similar motivation.

\subsection{Context}
Finally, the algorithms proposed here are designed for overweight, sedentary people and consequently, the space of exercise adaptation is within what may be considered safe for that population. Yet, sedentary people who are not overweight may find brisk walking or slight jogging useful. The range of exercise differs but the principle of adaption is general for both populations.

An important thing to note, however, is that our current algorithm design does not explicitly consider a trainee's motivation levels and other cognitive factors such as mood that may directly affect goal performance regardless of actual physical capability. This is an area where human coaches have a high degree of success because of their ability to interpret the trainee's emotional state and accordingly tailor the goal including motivating them appropriately.

Further, this aspect of monitoring and adjusting physical activity goals based on a trainee's motivation become very important in the context of coaching/rehabilitating individuals who have impaired abilities due to injuries and/or other health conditions. This is because these individuals/trainees may also have concomitant pain associated with physical activity and performing recommended exercises, which could adversely impact their motivation to perform them. In summary, our  algorithm would have to be extended in the future to account for other aspects of a trainee such as pain, motivation, mood etc. in addition to the physical exertion and affective attitude towards the exercise currently evaluated here in order to generalize to other populations and other coaching contexts.

Lastly, another contextual constraint for our current coaching algorithm design is that it requires a good method to quantify exercise intensity. Therefore, generalization to other clinical rehabilitation settings would require better a priori quantification of exercise intensity levels, which often vary highly between targeted exercises prescribed for individuals with impaired abilities. Therefore, future design modifications may have to include approaches to allow human coaches "in-the-loop" with the algorithm to adjust these parameters on a case by case basis for different trainees and over time.

\section{Conclusions}
Health behavior coaching is a challenging and complex problem for studying the design and analysis of interactive, intelligent agents. This domain puts modeling and reasoning about the human at the core of algorithm design. To be an effective coach, an intelligent agent must be long-living, continually personalize its coaching to each individual trainee, and gradually adapt interactions to the trainee's specific circumstance. Analysis of such long-living, interactive agents is challenging because their efficacy cannot be easily studied in laboratory settings, as is typical for AI research.

In this paper, we report how an interactive agent can be designed such that it has several properties desirable in a health coach. We propose a parameterized model for growth in aerobic capability which encodes factors that physical therapists use in their prescription of physical activity. We formulate an approach for \emph{adaptive goal setting} which uses the model for personal and temporal adaptation of aerobic goals. We implemented this approach in a smartphone application \textsc{NutriWalking}. To evaluate our approach, we conducted a four-pronged, task-centric paradigm to investigate different aspects of algorithm and system design. By simulating different kinds of trainees, we show that the algorithms adapt appropriately. Then, through an study with experts we show that the alogrithm produces goals that are difficult yet attainable, are safe, and are useful for the trainees in making progess towards the AHA goal. We conducted a usability study of \textsc{NutriWalking} with $15$ participants with diabestes or depression, who reported good acceptability. Finally, we deployed this application to $21$ participants with comorbid diabetes and depression for $6$ weeks. Our analysis shows that the coach helped participants increase the amount of exercise they do in a week. It did so by helping them set an appropriate goal to strive for every week. We evaluated three design decisions made during algorithm development. We argue that relying on self-reports for assessments can be problematic because people may overestimate their exercise capability. This optimism may negatively impact the trainee model and consequently, the coach can recommend exercise goals that are much harder than what the trainee chooses to perform. The current design allows the participant to make a choice about agreeing or disagreeing with the coach during the weekly goal setting. Our analysis revealed that some trainees are resistant to making their goal easier. This resistance translates to lower success rates, which may lead to trainees dropping out. Finally, our data suggest that rate of perceived exertion is a useful metric upon which to base exercise revisions. We show that this metric is predictive of a trainee's success or failure.

Promoting healthy behaviors (and curtailing harmful ones) is a challenging problem that can impact lives of millions around the world. Personal counselling has been show to be very effective in sustaining healthy behaviors. However, it is extremely resource-intensive and only available to few. Use of intelligent algorithms and human-computer interaction principles can help alleviate this problem by making personal adaptive counselling accessible to all. This paper takes an important step toward developing and deploying intelligent solutions for personal counselling.

Our research is a part of growing body of research on \emph{AI for social good} \citet{abebe2018mechanism} that aims to have beneficial impact on human society. Intelligent systems have been employed for allocation of security resources at ports \citep{shieh2012protect}, protecting biodiversity in conservation areas \citep{fang2016deploying}, and screening 800 million airport passengers annually throughout the USA \citep{brown2016one}. Research has also begun exploring intelligent algorithms and systems that can influence people's behavior to benefit society. Along with preventive healthcare, we have also explored how intelligent systems can influence people's transportation decisions to reduce energy expenditure \cite{mohan2019exploring, Mohan2019}. Often while designing AI systems, little attention is devoted to modeling the behavior of humans who invariably are crucial decision makers. Research on human-aware AI systems \citep{khampapati2018} seeks to address this gap and pose modeling humans as a question central to AI system design. Our paper contributes significantly to this research agenda by showing how models from behavioral psychology and exercise practitioners can be effectively incorporated in an AI system.

\section*{Acknowledgments}
The authors would like to thank Michael Silva for engineering \textsc{NutriWalking} and Peter Pirolli for contributions to the adaptive goal setting algorithm. The authors also acknowledge Dori Rosenberg, Evette Ludman, Paul Lozano, and James Ralston for protocol development and clinical oversight, as well as Catherine Lim, Leslie Jauregui and Ladia Albertson-Junkans for recruitment and data collection, and Kelly Ehrlich for project management. The work described in this paper was funded in part by Xerox corporation and by Kaiser Permanente Group Washington Research Institute. Additionally, the authors are grateful to Shekhar Mittal for advice on analyses presented in this paper.

\bibliographystyle{ACM-Reference-Format-Journals}
\bibliography{adaptive-coaching}
\end{document}